\documentclass[aps,prd,twocolumn,nofootinbib,preprintnumbers,superscriptaddress,showkeys,floatfix,amsmath,amssymb]{revtex4-1}

\newcommand{\bea}{\begin{eqnarray}}
\newcommand{\eea}{\end{eqnarray}}
\newcommand{\beq}{\begin{equation}}
\newcommand{\eeq}{\end{equation}}

\usepackage{mathrsfs}
\usepackage[utf8]{inputenc} 
\usepackage{slashed}

\usepackage{tikz,xcolor,hyperref}
\definecolor{lime}{HTML}{A6CE39}
\DeclareRobustCommand{\orcidicon}{
	\begin{tikzpicture}
	\draw[lime, fill=lime] (0,0) 
	circle [radius=0.16] 
	node[white] {{\fontfamily{qag}\selectfont \tiny ID}};
	\draw[white, fill=white] (-0.0625,0.095) 
	circle [radius=0.007];
	\end{tikzpicture}
	\hspace{-2mm}
}

\foreach \x in {A, ..., Z}{%
	\expandafter\xdef\csname orcid\x\endcsname{\noexpand\href{https://orcid.org/\csname orcidauthor\x\endcsname}{\noexpand\orcidicon}}
}

\newcommand{\orcid}[1]{\href{https://orcid.org/#1}{\textcolor[HTML]{A6CE39}{\aiOrcid}}}

\begin{document}

\title{Confining density functional approach to the QCD phase diagram at low temperatures and thermal twin stars  }

\author{David Blaschke\orcidA{}}
\email{david.blaschke@uwr.edu.pl}
\affiliation{
Institute of Theoretical Physics, University of Wroclaw, Max Born Pl. 9, 50-204, Wroclaw, Poland}
\affiliation{Helmholtz-Zentrum Dresden-Rossendorf (HZDR), Bautzner Landstrasse 400, 01328 Dresden, Germany}
\affiliation{Center for Advanced Systems Understanding (CASUS), Untermarkt 20, 02826 G\"orlitz, Germany}

\author{Oleksii Ivanytskyi\orcidB{}}
\email{oleksii.ivanytskyi@gmail.com}
\affiliation{
Institute of Theoretical Physics, University of Wroclaw, Max Born Pl. 9, 50-204, Wroclaw, Poland}
\affiliation{
Deutsches Zentrum f\"ur Astrophysik (DZA),
    Postplatz 1,
    02826 Görlitz, Germany
}

\date{\today}
\begin{abstract}
We present a density functional-based equation of state for warm, dense nuclear matter with a transition to deconfined quark matter for applications to simulations of supernova explosions and neutron star mergers, but also for the cosmological evolution of Q-balls.
For the quark matter equation of state, we employ a recently developed confining density functional approach while nuclear matter is described within a relativistic density functional model of the DD2 class. 
The phase transition is obtained by a Maxwell construction at constant entropy per baryon.
We discuss the solutions of TOV equations for isentropic hybrid stars 
for two hybrid equation of state models: DDf-SFM without and  DD2-$\chi$CDF with color superconductivity.
We find that at finite temperatures above a critical value of entropy per baryon sequences of disconnected third family branches ("thermal twin stars") may appear for the DDf-SFM model, while they are absent for the color superconducting model and at $T=0$.
We discuss the relation of this critical entropy per baryon to the Seidov criterion of gravitational instability for $T=0$ and find that it is a good guide.
We suggest that the presence of thermal twin stars may be regarded as an indicator for the core-collapse supernova explodability of massive blue supergiant stars with solar metallicity and thus serve as a new criterion for the reliability of hybrid equation of state models. 
By this argument, strong color superconductivity shall be excluded and it remains to be shown whether models with moderate diquark pairing could fulfill the thermal twin constraint.
For the case of symmetric matter, we compare the resulting hybrid EOS with the flow constraint by Danielewicz et al. and find a a sensitivity of the onset density for deconfinement to the presence or absence of color superconductivity. 
\end{abstract}

\date{\today}
\maketitle


\section{Introduction}

The investigation of the phase diagram of strongly interacting matter at low temperatures and high baryon densities is a challenge because it concerns a region where lattice quantum chromodynamics (QCD) and heavy-ion collision experiments from the high-temperature side and neutron stars at zero temperature cannot reach \cite{MUSES:2023hyz,Sorensen:2023zkk}. 
This white spot in the temperature-density plane is reserved for supernova explosions, neutron star mergers, and possibly, density fluctuations in the early Universe and their evolution in a Witten-type scenario of cosmic separation of phases \cite{Witten:1984rs,Gonin:2025uvc}.

The key question is whether in this region the transition between hadronic and quark matter may proceed as a first-order phase transition with the associated phenomena of a spinodal instability and the formation of so-called pasta phases \cite{Heiselberg:1992dx,Yasutake:2014oxa}. 
The occurrence of such spatio-temporal structures in the course of the dynamical evolution of these systems may leave observable signatures which in turn would provide evidence for the existence of a critical endpoint (CEP) of first-order phase transitions in the QCD phase diagram. Such a landmark is long sought-for because its existence and location would help to identify the universality class of QCD as the gauge-field theory of strong interactions.    

It was suggested in 2013 \cite{Blaschke:2013ana} that the very existence of a CEP in the QCD phase diagram would follow from the detection of a first-order deconfinement phase transition in neutron stars at vanishing temperature and high baryon density, since at high temperatures and vanishing baryon densities the results of lattice QCD simulations have evidence for a crossover transition with a pseudocritical temperature of 
$T_c=156.5\pm 0.15$ MeV \cite{Bazavov:2018mes}.  
In order to detect a strong deconfinement phase transition in neutron stars, the authors of \cite{Blaschke:2013ana} suggested looking for mass twin stars, i.e. pairs of pulsars with the same mass but significantly different radii. While the twin with the larger radius shall be a "normal" neutron star, the more compact twin should be the member of the third family
\cite{Gerlach:1968zz} of stable hybrid neutron stars, separated from the second family of neutron stars in the mass-radius diagram by a sequence of gravitationally unstable configurations. In the classification of hybrid stars that was introduced by Alford, Han and Prakash \cite{Alford:2013aca}
at the same time, the existence of such a disconnected family of hybrid stars would require a strong first-order phase transition with a jump in energy density $\Delta \varepsilon$ at the transition that would fulfill the Seidov criterion
\cite{1971SvA15347S},
\begin{equation}
    \label{eq:seidov}
    \Delta \varepsilon \ge (\varepsilon_c+3 p_c)/2~,
\end{equation}
where $\varepsilon_c$ and $p_c$ are the energy density and pressure at the onset of deconfinement, respectively.

While there is not yet observational evidence for the existence of cold mass twin neutron stars\footnote{In 2013, only two high-mass pulsars with a sufficiently precise measurement of an overlap mass range close to $2~M_\odot$ were known, PSR J1614-2230 with a mass of $1.97\pm 0.04~M_\odot$ \cite{Demorest:2010bx} and PSR J0348+0432 with $2.01\pm 0.04~M_\odot$ \cite{Antoniadis:2013pzd}.
It would have been a fantastic result and a confirmation of the existence of a CEP in the QCD phase diagram if the radii of these pulsars could have been measured with sufficient accuracy to conclude that they would be significantly different \cite{Benic:2014jia,Alvarez-Castillo:2016oln}!
Unfortunately, the radii of these pulsars could not be measured yet. 
However, there are different mass windows where twin stars could be detected, see the classification in \cite{Christian:2017jni}. 

With the neutron star interior composition explorer (NICER) experiment on the International Space Station, several simultaneous mass and radius measurements on millisecond pulsars could be performed. Among them are two pulsars in the same mass range with different, but rather uncertain radii
PSR J0614-3329 with 
$(R,M)=(10.29^{+1.01}_{-0.86}\,{\rm km},1.44^{+0.07}_{-0.06}~M_\odot)$ and PSR J0437-471 with 
$(R,M)=(11.9–15.5\,{\rm km},1.418\pm 0.044~M_\odot)$. 
Are these examples for the desperately sought-for twin stars?
It could be, but we don't know it yet since the radius measurements are still very uncertain.
},
there is another scenario which is of utmost relevance for the phenomenology related to the QCD phase diagram at low temperatures and high baryon densities. This concerns the possibility that at zero temperature the phase transition may have only a small latent heat, insufficient to fulfill the Seidov criterion \eqref{eq:seidov}, or even be a crossover\footnote{The latter possibility is called hadron-quark continuity and has been discussed, e.g., by Wetterich \cite{Wetterich:1999vd}, Sch\"afer and Wilczek \cite{Schafer:1998ef} and Hatsuda et al. \cite{Hatsuda:2006ps,Yamamoto:2007jn}.
Hatsuda et al. suggest that due to the mixing of the pairing gap in color superconducting quark matter and the quark mass gap in the scalar meson channel by the Fierz-transformed 't Hooft determinant interaction a coexistence of chiral and diquark gaps entails a crossover behavior which upon the melting of the diquark condensate goes over to a first-order phase transition, resulting in two critical endpoints in their low-temperature phase diagram. We will come back to the effects of the temperature dependent quark pairing in the main part of this work.}.

In a scenario without color superconductivity but with quark confinement at finite temperatures, the thermal excitations may increase the pressure in the quark phase stronger than in the hadronic one so that a reduction of the onset density of deconfinement accompanied with an increase in the latent heat can result. In such a case, the Seidov criterion may be fulfilled only at finite temperature, which leads to the phenomenon of thermal twin stars \cite{Hempel:2015vlg}. 
The existence of thermal twin stars may also be considered a precondition for the existence of a CEP.
For a recent discussion of this phenomenon, see \cite{Carlomagno:2023nrc,Carlomagno:2024vvr}.
The question arises and will be discussed in this work whether the thermal twin phenomenon is linked to the explodability of supernovae. From such a connection, if it exists, one could possibly derive constraints for the dense matter equation of state (EoS) and the structure of the QCD phase diagram. 

Already in 2011, a study of explodability of hybrid equations of state has been performed \cite{Fischer:2011zj}, where an exploding model based on the simple confining bag model EoS for quark matter \cite{Sagert:2008ka} was compared to an advanced model wherein quark matter was described microscopically by a three-flavor color superconducting chiral quark model of the NJL-type, albeit without confinement \cite{Blaschke:2005uj}.
In the latter model, due to strong color superconductivity the critical line for the onset of deconfinement in the low-temperature QCD phase diagram had an unconventional shape. It was bent towards higher densities when the temperature increases (see Fig. 1 of \cite{Blaschke:2010ka}), while the bag-model based non-superconducting hybrid EoS showed a lowering of the onset density with increasing temperature. In the supernova collapse accompanied with compression and heating of the protoneutron star core, the latter model could reach the onset of deconfinement fast enough, before mass accretion by fallback would lead to black-hole formation, and a successfully exploding supernova with a hybrid neutron star remnant results.
In the model \cite{Blaschke:2005uj}, while the protoneutron star core is heated up, the increasing density cannot catch up with the increasing onset density of the deconfinement. Therefore, the second shock at the phase border, which would trigger the explosion, cannot occur and the supernova fails.
As was shown by \cite{Hempel:2015vlg} for the bag model case, in the exploding scenario the thermal twin star formation was involved while increasing the onset (energy) density for the color superconducting NJL quark matter prevents the fulfillment of the Seidov criterion \eqref{eq:seidov} and no thermal twins are formed. 

Both hybrid EoS models of the study \cite{Fischer:2011zj} had serious caveats: the bag model based EoS could not describe massive neutron stars with $2~M_\odot$ and the NJL based EoS had no confinement which led to unphysical admixtures of unconfined quarks already at low temperatures 
$T\approx 50$ MeV.
The key to solving both problems was the development of a relativistic density functional approach within a Lagrangian formulation 
\cite{Kaltenborn:2017hus} that was based on an ansatz for the density-dependent interaction energy density functional motivated by the nonrelativistic string-flip model (SFM) \cite{Horowitz:1985tx} with the principle of color saturation within nearest-neighbor ranges as an effective color screening mechanism that allowed the quantum statistical treatment of confining interactions \cite{Ropke:1986qs}. 
With the confining density functional approach generalized to finite temperatures one could now show that quark deconfinement may serve as a supernova explosion mechanism for massive blue supergiant stars of 
$50~M_\odot$ \cite{Fischer:2017lag} and even $70~M_\odot$ \cite{Fischer:2021tvv} which with the conventional neutrino-driven explosion mechanism were determined to become black holes. 
The quark deconfinement supernova produces massive neutron stars of $~2\,M_\odot$ and a second neutrino burst as an observable signal. 

The aim of the present work is twofold.
First, we will introduce the field-theoretic approach to the confining density functional (CDF) for quark matter in Section \ref{sec_model} and discuss in Section \ref{sec_results} the QCD phase diagram that results from this approach on the SFM basis as well as for its generalization to a chirally symmetric ansatz with diquark current-current interactions for addressing the effects of color superconductivity \cite{Ivanytskyi:2022oxv}, the $\chi$CDF model. 
For the first time, we will contrast the cases of isospin-symmetric matter and neutron star matter and highlight the qualitatively new features that arise from strong color superconductivity.
We will show the trajectories of constant entropy per baryon for our model EoS, which are guiding the understanding of dynamical evolution of heavy-ion collisions, supernovae and the hadronising quark-plasma in an inhomogeneous big bang model of cosmology since apart from dissipative effects the hydrodynamic evolution is governed by the conservation of entropy and (net) particle number\footnote{For examples of simulations with a QCD phase transition which produce such trajectories, see \cite{Fischer:2017lag,Fischer:2021tvv} for core-collapse supernovae (CCSN) and \cite{Huovinen:2007xh} for heavy-ion collisions.}. 
Throughout this work we will consider nonstrange, two-flavor hadronic and quark matter. Leptons are added when electric neutrality and $\beta$- equilibrium are required for neutron star matter.  Under this term we understand here the state of matter that is found in late-stage, deleptonized, neutrino-transparent compact stars.

Second, we argue that thermal twin stars as a property of hybrid EoS models may be indicators for the CCSN explodability of massive blue supergiant progenitor stars with solar metallicity. Since with SN2005gl \cite{2009Natur.458..865G}, there is a prominent example for an observation of the CCSN explosion of a $M\gtrsim 50 \,M_\odot$ progenitor star, this explodability may be considered as an observational fact that cannot be explained with the standard neutrino-driven explosion mechanism. 
As long as the QCD deconfinement mechanism suggested in \cite{Fischer:2017lag} is the only mechanism to explain the explodability in cases like SN2005gl, we want to promote here the conjecture of a thermal twin - CCSN explodability connection raised already in \cite{Hempel:2015vlg} to the {\it thermal twin constraint:
\\
A relevant EOS for the QCD phase diagram has to have the thermal twin property because only in this way the explodability of massive blue supergiant progenitor stars as an observed phenomenon can be explained.}

It is a striking result of this work that the $\chi$CDF model for a parametrization motivated by multimessenger observations of neutron stars does not yield an EoS with thermal twin stars! 
It corresponds to a QCD phase diagram where the hadronic phase is separated from the normal quark matter phase by a corridor of color superconducting matter. 
Such a phase structure is reported here for the first time.
In Section \ref{sec_summary} we summarize the main findings and discuss progress achieved in this work in the context of other approaches. 
Section \ref{sec_conclusion} gives our conclusions and an outlook to planned future developments of the presented approach. 


\section{Confining Density Functional Approach for Dense QCD Matter}
\label{sec_model}

The phenomenological mechanism of mimicking color confinement of QCD, which is adopted within the density functional approaches considered in this work, corresponds to fast growth of the mean-field self-energy of quarks in the confining region.
This makes their excitations energetically suppressed, and consequently excludes quarks from thermodynamically relevant degrees of freedom. 
Microscopically this mechanism can be interpreted as a result of large masses of quarks, which makes hadron decays kinematically forbidden in the confining region.

In this section, we give a general formulation of the CDF approach to two-flavor quark matter first developed in \cite{Kaltenborn:2017hus}.
Later in \cite{Ivanytskyi:2022oxv} it was formulated in a chirally symmetric form, extended to the next order beyond the mean-field approximation, and generalized to the CS case.
The corresponding Lagrangian can be written as
\begin{eqnarray}
\label{I}
   \mathcal{L}=\overline{q}(i\slashed\partial-\hat m+\hat\mu\gamma_0)q-\mathcal{U}_\chi-\mathcal{U}_V-\mathcal{U}_I-\mathcal{U}_D.
\end{eqnarray}
It includes two-flavor quark spinor $q^{T}=(u,d)$ and diagonal matrices of quark masses $\hat{m}={\rm diag}(m_u,m_d)$ and chemical potentials $\hat{\mu}={\rm diag}(\mu_u,\mu_d)$ acting in the flavor space.
In the following, we neglect the mass splitting of the $u$ and $d$ quarks so that $m_u=m_d=m$.
The chemical potentials of quark flavor $f$ are expressed through the chemical potentials of baryon charges $\mu_B$ and electric charges $\mu_Q$ as $\mu_f=\mu_B/3+\mu_QQ_f$, where $Q_f$ is the electric charge. 

It is worth mentioning that in the general case CS leads to color inbalance of quark matter \cite{Baym:2017whm}. Restoring it, i.e. providing color neutrality of quark matter, requires introducing two color chemical potentials $\mu_3$ and $\mu_8$ associated to the diagonal Gell-Mann matrices $\lambda_3$ and $\lambda_8$ of the color group \cite{Powell:2013cq}.
As a results the Lagrangian (\ref{I}) should contain $\hat\mu+\mu_3\lambda_3+\mu_8\lambda_8$ rather than just $\hat\mu$.
While $\mu_3$ vanishes in the color basis adopted below, $\mu_8$ typically attains values, which are small compared to the values of chemical potentials of quarks.
Therefore, in this work for the sake of simplicity we neglect $\mu_8$ since it does not lead to significant modification of EoS of quark matter.

Within chiral quark models of the Nambu-Jona-Lasinio (NJL) type, the chiral dynamics of QCD is modeled by connecting the dynamically generated mass gap to chiral condensate $\langle\overline{q}q\rangle$ \cite{Scarpettini:2003fj,Blaschke:2007np,Hell:2009by,Ivanytskyi:2024zip}.
Therefore, if formulated in a chirally symmetric form, the confining potential in Eq. (\ref{I}), i.e. $\mathcal{U}_\chi$, is responsible for spontaneous breaking and dynamical restoration of chiral symmetry of QCD.
In this case $\mathcal{U}_\chi$ should depend on the quark operators $\mathcal{O}_{\chi,0}=\overline{q}q$ and $\mathcal{O}_{\chi,l}=\overline{q}i\gamma_5\tau_lq$ with $\tau_l$ being flavor Pauli matrices (see \cite{Ivanytskyi:2022oxv} for details).
The independence of $\mathcal{U}_\chi$ on $\overline{q}i\gamma_5\vec{\tau}q$ is equivalent to treating the chiral dynamics of QCD implicitly as was done in \cite{Kaltenborn:2017hus}, where additional medium dependence of the confining interaction was introduced by defining $\mathcal{O}_{\chi,4}=q^+ q$.

Nonperturbative gluon exchange among quarks generates a strong repulsive interaction in dense quark matter \cite{Song:2019qoh}. 
This repulsive interaction is important for reaching the two solar mass limit of NSs \cite{Baym:2017whm}.
In Eq. (\ref{I}) this is taken into account by the vector interaction potential $\mathcal{U}_V$ that depends on the quark operators $\mathcal{O}_{V,\mu}=\overline{q}\gamma_\mu q$.

Similarly to the isospin-sensitive interaction in nuclear matter that determines the nuclear symmetry energy \cite{Baldo:2016jhp}, interactions in dense quark matter are sensitive to quark flavors \cite{Baym:2017whm}.
Within the CDF this is modeled by the isospin-sensitive vector interaction potential $\mathcal{U}_I$.
It depends on the quark operators $\mathcal{O}_{I,\mu l}=\overline{q}\gamma_\mu \tau_l q$.

At sufficiently high density quark matter experiences a Cooper instability \cite{Cooper:1956zz}, which at two flavors leads to formation of a two flavor color-superconductivity (2SC). 
A 2SC state of quark matter is constituted by the Bose-Einstein condensate of diquarks \cite{Buballa:2003qv} and is important for the NS phenomenology \cite{Alford:2007xm,Baym:2017whm,Kojo:2020ztt,Ivanytskyi:2022oxv,Ivanytskyi:2022wln,Ivanytskyi:2022bjc,Gartlein:2023vif,Gartlein:2024cbj,Gholami:2024ety,Gartlein:2025zhd,Christian:2025dhe,Sabatucci:2026qcz}.
This phenomenon is included in CDF via the diquark pairing potential $\mathcal{U}_D$ that depends on the quark operator $\mathcal{O}_{D,0}=\overline{q}i\gamma_5\tau_2\lambda_2q^c$, $\mathcal{O}_{D,1}=\overline{q}i\gamma_5\tau_2\lambda_5q^c$, $\mathcal{O}_{D,2}=\overline{q}i\gamma_5\tau_2\lambda_7q^c$ and their Hermitian conjugates $\mathcal{O}_{D,a\le2}^+\equiv\mathcal{O}_{D,a+3}$.
Here $\lambda_a=2,5,7$ are antisymmetric color Gell-Mann matrices and $q^c=i\gamma_2\gamma_0{\overline{q}}^T$ stands for charge conjugated quark field.

The potentials $\mathcal{U}_X$ from Eq. (\ref{I}) can be expanded around the mean-field solution, i.e. around the expectation values $\langle \vec{\mathcal{O}}_X\rangle$ of the corresponding operators.
Hereafter $X=\chi,V,I,D$ and the vector $\vec{\mathcal{O}}_X$ is composed of the components $\mathcal{O}_{X,{i_X}}$ that are labeled by the subscript index $i_X$.
It is worth mentioning that in the general case the number of components of the vectors $\vec{\mathcal{O}}_X$ is not the same.
With these notations the second order expansion of the interaction potentials can be written as
\begin{eqnarray}
    \mathcal{U}_X\simeq\mathcal{U}_X^{MF}&+&
    \left(\vec{\mathcal{O}}_X-\langle \vec{\mathcal{O}}_X\rangle\right)
    \vec{\Sigma}_X\nonumber\\
    \label{II}
    &-&
    \left(\vec{\mathcal{O}}_X-\langle \vec{\mathcal{O}}_X\rangle\right)
    \hat{G}_X
    \left(\vec{\mathcal{O}}_X-\langle \vec{\mathcal{O}}_X\rangle\right).
\end{eqnarray}
The first term in this expression stands for the interaction potential $\mathcal{U}_X$ evaluated at mean field, i.e. when the corresponding quark operators $\vec{\mathcal{O}}_X$ are replaced by their expectation values.
The next correction represents the first order term, which includes the vector of expansion coefficients
\begin{eqnarray}
    \label{III}
    \vec{\Sigma}_X=\frac{\partial \mathcal{U}_X^{MF}}{\partial \langle\vec{\mathcal{O}}_X\rangle}.
\end{eqnarray}
Its components are nothing but the mean-field self energies of quarks generated by the corresponding interaction potentials. 
The second order term in the expression (\ref{II}) is given in terms of the matrix of effective medium-dependent couplings of the corresponding interaction channels
\begin{eqnarray}
    \label{IV}
    \hat{G}_X=-\frac{1}{2}\frac{\partial }{\partial \langle\vec{\mathcal{O}}_X\rangle}\otimes\vec{\Sigma}_X,
\end{eqnarray}
where $\otimes$ denotes the direct product of vectors.

At the mean-field level the only nonvanishing self energies are the scalar-isoscalar $\Sigma_{\chi,1}\equiv\Sigma_S$, vector-isoscalar $\Sigma_{V,0}\equiv\Sigma_V$, vector-isovector $\Sigma_{I,03}\equiv\Sigma_I$ and and diquark $\Sigma_{D,0}\equiv\Sigma_D$ ones.
The first three of them renormalize quark masses and chemical potentials leading to the effective ones 
\begin{eqnarray}
    \label{V}
    m^*&=&m+\Sigma_S,\\
    \label{VI}
    \hat{\mu}^*&=&\hat\mu-\Sigma_V-\tau_3\Sigma_I.
\end{eqnarray}
The modulus of the diquark self energy defines the energy gap of quarks due to their pairing, i.e. $\Sigma_D\Sigma_D^*=\Delta^2$.
It enters the inverse Nambu-Gorkov propagator along with the effective quark mass and chemical potentials. 
At the mean-field level it reads
\begin{eqnarray}
   \label{VII}
   \hat{\mathcal{S}}^{-1}=
   \left(
   \begin{array}{l}
   i\slashed\partial-m^*+\gamma_0\hat\mu^*\hspace*{.1cm}
   \hspace*{.5cm}i\gamma_5\tau_2\lambda_2\Sigma_D\\
   \hspace{.4cm}i\gamma_5\tau_2\lambda_2\Sigma_D^*
   \hspace*{.6cm}i\slashed\partial-m^*-\gamma_0\hat\mu^*
   \end{array}
   \right).
\end{eqnarray}
The corresponding Nambu-Gorkov quark field is defined as $\mathcal{Q}^T=(q,q^c)/\sqrt{2}$.
The presence of the second Gell-Mann color matrix in this expression signals that only red and green quark color states of the 2SC quark matter are paired in the chosen color basis (see Ref. \cite{Buballa:2003qv} for details).
To label the color states below we use the subscript index ``$c$''.
Therefore, the color structure of the pairing gap is $\Delta_c=(\Delta,\Delta,0)$.

The eigen values of the quark propagator in the momentum representation give single particle energies of quarks (superscript index ``$+$'') and antiquarks (superscript index ``$-$'') shifter by the chemical potentials of the corresponding flavors 
\begin{eqnarray}
    \label{VIII}
    \epsilon_{fc{\bf k}}^a={\rm sgn}(\epsilon_{\bf k}-a\mu_f^*)
    \sqrt{(\epsilon_{\bf k}-a\mu_f^*)^2+\Delta_c^2},
\end{eqnarray}
where $\bf k$ is three momentum and $\epsilon_{\bf k}=\sqrt{{\bf k}^2+{m^*}^2}$.

Using the above definitions along with the Nambu-Gorkov propagator of quarks, the Lagrangian of an ordinary CDF can be brought to the form of current-current interaction of the NJL model.
It is worth mentioning that the NJL model itself is a density functional approach to quark matter.
This form of the CDF approach allows treating it via the standard procedure of Hubbard-Stratonovich bosonization. 
The latter introduces the vector of collective bosonic fields $\vec{\phi}_X$, coupled to the corresponding $\vec{\mathcal{O}}_X-\langle\vec{\mathcal{O}}_X\rangle$.
Formally, this procedure can be presented via the identity
\begin{eqnarray}
    &&\exp\left[\int dx
    \left(\vec{\mathcal O}-\langle \vec{\mathcal{O}}\rangle\right)
    \hat{G}
     \left(\vec{\mathcal O}-\langle \vec{\mathcal{O}}\rangle\right)\right]\nonumber\\
    \label{IX}
    &&=\int[\mathcal{D}\vec{\phi}]
    \exp\left[\int dx\left(
    \vec{\phi}\left(\vec{\mathcal{O}}-\langle \vec{\mathcal{O}}\rangle\right)-\frac{\vec{\phi}\hat{G}^{-1}\vec{\phi}}{4}\right)\right],\,
\end{eqnarray}
where the subscript index ``$X$'' is suppressed for shortening the notations.
In this relation $\int dx$ stands for the integration over the spacial volume $V\equiv\int d{\bf x}$ and temporal component ranging from $0$ to inverse temperature $\beta\equiv1/T$.
It is worth mentioning that the field $\vec{\phi}_\chi$ is related to scalar $\sigma$-meson and pseudoscalar pions, and $\vec{\phi}_V$ and $\vec{\phi}_I$ represent the vector-isoscalar $\omega$-meson and vector-isovector $\rho$-mesons.
The field $\vec{\phi}_D$ and its complex conjugate represent diquarks and antidiquarks, respectively.

The above notations allow writing the bosonized Lagrangian of the CDF approach as
\begin{eqnarray}
    \label{X}
    \mathcal{L}^{\rm bos}&=&\overline{\mathcal{Q}}\hat{\mathcal{S}}^{-1}\mathcal{Q}-
    \sum_X\left(\mathcal{U}_X-\vec{\Sigma}_X\langle\vec{\mathcal O}_X\rangle\right)\nonumber\\
    \label{XI}
    &+&\sum_X
    \left(
    \vec{\phi}_X\left(\vec{\mathcal O}_X-\langle \vec{\mathcal O}_X\rangle\right)-\frac{\vec{\phi}_X\hat{G}_X^{-1}\vec{\phi}_X}{4}\right).
\end{eqnarray}
From this we obtain the Euler-Lagrange equations for the auxiliary bosonic fields as
\begin{eqnarray}
    \label{XI}
    \vec{\phi}_X=2\hat{G}_X\left(\vec{\mathcal{O}}_X-\langle \vec{\mathcal{O}}_X\rangle\right).
\end{eqnarray}
It follows from these equations that $\vec{\phi}_X$ vanishes at the mean-field approximation, since in this case $\vec{\mathcal{O}}_X$ is substituted by its expectation value.
Thus, the auxiliary bosonic fields acquire beyon mean-field correlations of (anti-)quarks, i.e. mesons and diquarks. 
At temperatures and baryon densities typical for interiors of thermal NSs, the contribution of these correlations is subdominant.
Therefore, following \cite{Kaltenborn:2017hus,Ivanytskyi:2022oxv,Ivanytskyi:2022wln} in this work we restrict the consideration of EoS to the mean-field level, i.e. neglect the last terms in Eq. (\ref{X}).
However, these terms are important for connecting the mean-field EoS to the vacuum mass spectrum of mesons, i.e. to the low energy phenomenology of QCD as has been done in \cite{Ivanytskyi:2022oxv}.

Neglecting the second order terms in the bosonized Lagrangian makes it quadratic in quark fields, which can be integrated out to obtain the thermodynamic potential
\begin{eqnarray}
    \label{XII}
    \Omega=-\frac{1}{2\beta V}{\rm tr}\ln\left(\beta\mathcal{S}^{-1}\right)+
    \sum_X\left(\mathcal{U}_X-\vec{\Sigma}_X\langle\vec{\mathcal O}_X\rangle\right).
\end{eqnarray}
The factor $1/2$ in the first term of this expression compensates the artificial doubling of degrees of freedom in the Nambu-Gorkov formalism, while the trace is carried over the Nambu-Gorkov,  Dirac, flavor, color, three-momentum, and fermion Matsubara indexes.
Performing this operation, one gets
\begin{eqnarray}
    \Omega&=&-2\sum_{fca}\int\frac{d{\bf k}}{(2\pi)^3}
    \left[\frac{g_{\bf k}\epsilon_{fc{\bf k}}^a}{2}
    +T\ln\left(1+e^{-\beta\epsilon_{fc{\bf k}}^a}\right)\right]
    \nonumber\\
    \label{XIII}
    &+&\sum_X\left(\mathcal{U}_X-\vec{\Sigma}_X\langle\vec{\mathcal O}_X\rangle\right).
\end{eqnarray}
Here $g_{\bf k}$ is the regulator of the zero point term.
Below this regulator is specified for each of the considered versions of the CDF.
It provides the convergence of the momentum integral in the zero point term of $\Omega$ and all the thermodynamic quantities related to its derivatives such as baryon and entropy densities, and self energies of quarks $\vec{\Sigma}_X$.
At the mean-field level, the latter should be properly adjusted so that the thermodynamic potential is minimized.
Thus, the self energies are defined by the conditions
\begin{eqnarray}
    \label{XIV}
    \frac{\partial\Omega}{\partial\vec{\Sigma}_X}=0.
\end{eqnarray}
These requirements ensure thermodynamic consistency of the model and allow self-consistent determination of the expectation values of the quark operator $\langle\vec{\mathcal{O}}_X\rangle$, which are required to find $\Omega$.
The latter can be used to obtain pressure as the negative of the thermodynamic potential shifted by its vacuum value $\Omega_{\rm vac}$, i.e. $p=\Omega_{\rm vac}-\Omega$.
The densities of baryon and electric charges, entropy, and energy are found as $n_B=\partial p/\partial\mu_B$, $n_Q=\partial p/\partial\mu_Q$, $s=\partial p/\partial T$, $\varepsilon=\mu_B n_B+Ts-p$, respectively.
Another important characteristic of quark matter is chiral condensate $\langle\overline{q}q\rangle=\partial\Omega/\partial m$.
The quantities mentioned constitute the EoS of a CDF approach that is specified by a particular choice of the interaction potentials $\mathcal{U}_X$. 

\subsection{CDF of the string-flip model of quark matter}
\label{SFM}

The CDF approach to quark matter from \cite{Kaltenborn:2017hus} is based on the String Flip Model (SFM) \cite{Horowitz:1985tx,Ropke:1986qs}.
Within the SFM the QCD vacuum is a dual color superconductor, which expels gluon fields just as electromagnetic superconductor expels magnetic field.
As a result, tension lines of the gluon fields generated by confined quarks are collimated into narrow elongated structures extending between the quarks, confining gluon strings.
The energy of these strings is proportional to their length, i.e. to the mean separation between quarks.
Therefore, in-medium self energy of quarks should scale inverse proportionally to the cubic root of their number density. 
In the confining region this number density is very close to the scalar density.
Within the no-sea approximation adopted in \cite{Kaltenborn:2017hus} the zero point term in the thermodynamic potential is absent ($g_{\bf k}=0$) and the scalar density coincides with the negative of chiral condensate. 
Thus, within the SFM $\Sigma_\chi\propto \langle\overline{q}q\rangle^{-1/3}$.
Such behavior of the scalar self energy of quarks can be provided by the interaction potential
\begin{eqnarray}
    \label{XV}
    \mathcal{U}_\chi=-D_0\exp\left(-v^2\vec{\mathcal{O}}_V^2\right)(\overline{q}q)^{2/3},
\end{eqnarray}
where the coupling $D_0$ has the meaning of string tension and $v$ is the generalized excluded volume parameter (see Ref. \cite{Kaltenborn:2017hus} for details).
The ``$-$'' sign in the right hand side of this expression guaranties the attractive character of the potential $\mathcal{U}_\chi$.

The vector-isoscalar interaction potential is defined as 
\begin{eqnarray}
    \label{XVI}
    \mathcal{U}_V=a~\vec{\mathcal{O}}_V^2+
    \frac{b~\vec{\mathcal{O}}_V^4}{1+c~\vec{\mathcal{O}}_V^2}.
\end{eqnarray}
The parameters $a$, $b$ and $c$ are specified below.
The second term in this expression is introduced to improve the behavior of the model at high densities.

The vector-isovector potential has been introduced to the present model in \cite{Fischer:2017lag}. 
It has the simple quadratic form
\begin{eqnarray}
    \label{XVII}
    \mathcal{U}_I=d~\vec{\mathcal{O}}_I^2.
\end{eqnarray}
The strength of the isospin-sensitive interaction given by this potential is controlled by the coupling $d$.
In \cite{Fischer:2017lag} it was adjusted to provide continuity of the symmetry energy across the deconfinement phase transition.

The SFM ignores the possibility of 2SC in NSs.
Its diquark potential vanishes, i.e. $\mathcal{U}_D=0$ and $\vec{\Sigma}_D=0$.
The corresponding single particle energy $\epsilon_{fc{\bf k}}^a=\epsilon_{\bf k}-a\mu_f^*$ is obtained by setting $\Delta=0$ in Eq. (\ref{VIII}).

\begin{table}[t]
\centering
\begin{tabular}{|c|c|c|c|c|c|c|}
\hline    
$m$  &  $\sqrt{D_0}$     &     $v$      &         $a$      &       $b$        &      $c$     &      $d$         \\
$[\rm MeV]$&$[\rm MeV]$&$[\rm fm^3]$ & $[\rm MeV~fm^3]$ & $[\rm MeV~fm^9]$&$[\rm fm^6]$&$[\rm MeV~fm^3]$ \\ \hline
 5.5  &   265    &  0.624   &   -4  & 1.6  &   0.025  &   80          \\ \hline
\end{tabular}
\caption{Parameters of the SFM used in this work.} 
\label{table1}
\end{table}
In this work, we use the set of parameters of the SFM from  \cite{Fischer:2017zcr}.
The corresponding values are given in Table \ref{table1}.

\subsection{Chiral CDF for color superconducting quark matter}
\label{CCDF}

The interaction potential $\mathcal{U}_\chi$ of the SFM does not respect the chiral symmetry, which is a property of QCD interactions. 
This does not allow the SFM to address the low energy phenomenology of QCD described by the Goldstone theorem \cite{Goldstone:1962es} and the Gell-Mann-Oakes-Renner relation \cite{Gell-Mann:1968hlm}.
To remove this issue in \cite{Ivanytskyi:2022oxv} the CDF was formulated in a chirally symmetric form, which is referred to as $\chi$CDF.
For this, the argument of the potential $\mathcal{U}_\chi$ was chosen in chirally symmetric form as
\begin{eqnarray}
    \label{XVIII}
    \mathcal{U}_\chi=D_0\left[(1-\alpha)\langle\vec{\mathcal{O}}_\chi\rangle_{\rm vac}^2-\vec{\mathcal{O}}_\chi^2\right]^{1/3}.
\end{eqnarray}
Similarly to SFM, the string tension parameter $D_0$ in this expression defines the strength of the interaction, the vacuum value $\langle\vec{\mathcal{O}}_\chi\rangle_{\rm vac}^2=\langle\overline{q}q\rangle_{\rm vac}^2$ is introduced for convenience, and the parameter $\alpha$ controls the vacuum value of the effective quark mass $m^*_{\rm vac}$.
The latter scales as $m^*_{\rm vac}\propto\alpha^{-2/3}$ and, consequently, diverges at $\alpha=0$ (see Ref. \cite{Ivanytskyi:2022oxv} for details).  
At $\alpha=1$ and $\vec{\mathcal{O}}_\chi$ replaced by $\overline{q}q$ the potential $\mathcal{U}_\chi$ of the present model coincides with that of the SFM if the generalized excluded volume parameter is set zero, i.e., $v=0$. 
However, this coincidence is formal since, unlike the SFM, the present model goes beyond the no-sea approximation and accounts for the zero point contribution.
The corresponding regulator has the Gaussian form
\begin{eqnarray}
    \label{XIX}
    g_{\bf k}=\exp\left(-{\bf k}^2/\Lambda^2\right)
\end{eqnarray}
with $\Lambda$ being a soft momentum cutoff.

The vector-isoscalar and vector-isovector interaction potentials read
\begin{eqnarray}
\label{XX}
\mathcal{U}_{V,I}=
G_{V,I}^{\rm vac}\int_0^{\vec{\mathcal{O}}_{V,I}^2} d\zeta~\left[1+\frac{8}{9M_g^2}\left(\frac{\pi^4\zeta}{4}\right)^{1/3}\right]^{-1},
\end{eqnarray}
where $G_{V,I}^{\rm vac}$ is the corresponding coupling in vacuum. 
The chosen parameterization of $\mathcal{U}_{V,I}$ is motivated by the analysis of in-medium nonperturbative one-gluon exchange between quarks \cite{Song:2019qoh}.
The main parameter of this process is the nonperturbative gluon mass $M_g$. 
The Fock energy density generated by such exchange process can be written in the typical mean-field form $\varepsilon_{V,I}=G_{V,I}^{\rm eff}\langle\vec{\mathcal O}_{V,I}\rangle^2$ with the effective coupling
\begin{eqnarray}
    \label{XXI}
    G_{V,I}^{\rm eff}=\frac{G_{V,I}^{\rm vac}}{1+\frac{8}{9M_g^2}\left(\frac{\pi^4\langle\vec{\mathcal O}_{V,I}\rangle^2}{4}\right)^{1/3}},
\end{eqnarray}
which for the vector-isoscalar interaction was suggested in \cite{Song:2019qoh}.
In \cite{Ivanytskyi:2022bjc} the potential $\mathcal{U}_V$ was formulated to ensure $\Sigma_V=2G_V^{\rm eff}\langle q^+q\rangle$.
In this work, we extend the parameterization (\ref{XXI}) to vector-isovector interaction as well.
Introducing the vector-isovector interaction channel to the $\chi$CDF is a new element of the present work. 

In \cite{Ivanytskyi:2022bjc} the diquark pairing interaction was parameterized similarly to the vector repulsion among quarks. 
Thus 
\begin{eqnarray}
\label{XXII}
\mathcal{U}_D=
G_D^{\rm vac}\int_0^{\vec{\mathcal{O}}_D\hat{\mathcal{A}}\vec{\mathcal{O}}_D} d\zeta~\left[1+\frac{8}{9M_g^2}\left(\frac{\pi^4\zeta}{4}\right)^{1/3}\right]^{-1},
\end{eqnarray}
where $\hat{\mathcal{A}}$ is andtidiagonal matrix with unit elements that acts in the space of components of the vector $\vec{\mathcal{O}}_D$ and $G_D^{\rm vac}$ is the diquark coupling in vacuum.
The latter should be chosen so that the vacuum state is stable against the formation of color superconductivity (see \cite{Ivanytskyi:2022oxv} for details).

As at high densities the expectation values of the operators $\vec{\mathcal{O}}_{V,I,D}$ are large. As a result, the integrals in Eqs. (\ref{XX}) and (\ref{XXII}) come mostly from large $\zeta$ so that the integrand can be approximated as $\propto\zeta^{-1/3}$.
This leads to $\mathcal{U}_{V,I}\propto(\vec{\mathcal{O}}_{V,I}^2)^{2/3}$ and $\mathcal{U}_{D}\propto(\vec{\mathcal{O}}_D\hat{\mathcal{A}}\vec{\mathcal{O}}_D)^{2/3}$ at high densities.
This scaling entails that the conformal limit of QCD is reached within the CDF \cite{Ivanytskyi:2022bjc}.

The $\chi$CDF presented in this work has seven parameters, they are $m$, $D_0$, $\alpha$, $\Lambda$, $G_V^{\rm vac}$, $G_I^{\rm vac}$ and $G_D^{\rm vac}$. 
Most of them are fixed based on the well established strategy of chiral quark models, i.e. by addressing the vacuum mass spectrum of hadrons.
The corresponding masses are extracted from the propagators obtained within the Gaussian approximation \cite{Ivanytskyi:2022oxv}\footnote{The same mesonic propagators can be also obtained via resummation of the infinite series of quark-antiquark scattering diagrams generated by the one-loop polarization operators within the random phase approximation \cite{Klevansky:1992qe}.}.
The parameters of the pseudoscalar interaction were fixed according to the mass $M_\pi=140$ MeV and the decay constant $F_\pi=90$ MeV of pions, the mass of the narrowest scalar meson $f_0(980)$, i.e., $M_\sigma=980$ MeV and the pseudocritical temperature of chiral crossover $T_{PC}=163$ MeV determined by the position of the peak of chiral susceptibility (see Ref. \cite{Ivanytskyi:2022oxv} for details).
The masses of vector-isoscalar $\omega$-meson ($M_\omega=783$ MeV) and vector-isovector $\rho$ -meson ($M_\rho=775$ MeV) were used to fix the parameters of the couplings $G_V^{\rm vac}$ and $G_I^{\rm vac}$, respectively.
These values are related to the vacuum value of scalar coupling as $G_V^{\rm vac}/G_S^{\rm vac}=0.452$ and $G_V^{\rm vac}/G_S^{\rm vac}=0.455$, which is very close to the value $0.5$ based
on a Fierz transformation of the (massive) vector boson exchange \cite{Buballa:2003qv}. 
This allows us to assume that the value of the diquark coupling should also be close to the one predicted by the Fierz transformation argument, i.e. $G_D^{\rm vac}/G_S^{\rm vac}=0.75$
The used value of the nonperturbative gluon mass provides good agreement with the observational data on NSs \cite{Ivanytskyi:2022bjc,Gartlein:2023vif,Gartlein:2024cbj}.
The values of the model parameters fixed as described above are given in Table \ref{table2}.
\begin{table}[t]
\centering
\begin{tabular}{|c|c|c|c|c|c|c|c|}
\hline    
$m$  &  $\Lambda$ & $\alpha$ & $D_0/\Lambda^2$ & $G_V^{\rm vac}\Lambda^2$ & $G_I^{\rm vac}\Lambda^2$ & $G_D^{\rm vac}\Lambda^2$&$M_g/\Lambda$\\{}
[MeV]&[MeV] & & &  &  &  & \\ \hline
  4.2 & 573 & 1.43  & 1.39 & 2.69  & 2.71 & 4.04 & 1.05\\ \hline
\end{tabular}
\caption{Parameters of the $\chi$CDF used in this work.} 
\label{table2}
\end{table}
%

\section{Results for the thermodynamics and QCD phase diagram}
\label{sec_results}

The CDFs presented in Sec. \ref{sec_model} provide reliable description of quark matter that exists at high densities.
At small densities quarks remain confined and hadrons are relevant degrees of freedom of QCD. 
Although SFM and $\chi$CDF provide a systematic approach to a thermodynamically consistent treatment of hadrons within the generalized Beth-Uhlenbeck approach (see e.g. \cite{Blaschke:2023pqd} for details), in this work we follow a simpler strategy, which is common for modeling NSs. 
For this we construct a hybrid quark-hadron EoS via matching its high density quark matter part described within a CDF and low density hadron part by means of the Maxwell construction\footnote{
In the case of multiple charges a so-called Glendenning construction of the phase transition has been introduced on the basis of the Gibbs conditions for phase equilibria \cite{Glendenning:1992vb}. 
However, this construction neglects the important effects of surface tension which lead to the formation of structures in the mixed phase (so-called "pasta phases") which call for the inclusion of Coulomb correlations and charge screening. A detailed investigation of a 
mixed-phase model with quark-hadron pasta calculations at T=0 
\cite{Maslov:2018ghi} has shown that for realistic values of the surface tension the difference between the full pasta phase calculation and a Maxwell construction is minor and restricted to just a "rounding of edges", see also \cite{Voskresensky:2002hu} and references therein. It has been shown in \cite{Ayriyan:2017nby}
that this rounding of edges does not destroy the twin star property for a decent range of parameter values. We expect that similar conclusions could be drawn when a finite-temperature extension of the pasta phase description would be applied so that the critical value $s/n_B=1.75$ for he occurrence of thermal twins could remain largely unaffected.
}.
Within this approach the position of deconfinement phase transition is defined by the Gibbs conditions of phase equilibrium, namely that for a given temperature the pressure and chemical potentials of the quark and hadronic matter phases coincide.  

In the case of isospin symmetric matter the electric chemical potential should be set equal to zero what provides equal numbers of isospin partners, i.e. neutrons and protons in hadron phase and $u$- and $d$-quarks in quark phase.
Electric neutrality of the NSs matter is provided by the proper amount of electrons.
In the NS matter case, the condition of $\beta$-equilibrium is $\mu_n=\mu_p+\mu_e$ in the hadron phase
\footnote{Similarly to quarks, chemical potentials of neutrons and protons are defined as $\mu_{n,p}=\mu_B+\mu_Q Q_{n,p}$, where $Q_{n,p}$ is the corresponding electric charge.} 
and $\mu_d=\mu_u+\mu_e$ in the quark phase, allows us to express the chemical potential of electrons as $\mu_e=-\mu_Q$.
Thus, the electric chemical potential determines the number of electrons in NSs.
This amount should be adjusted to provide electric neutrality of the NS matter, which excludes $\mu_Q$ from the list of independent thermodynamic quantities.
In this work, we treat electrons as noninteracting spin-$1/2$ fermions with mass $m_e=0.511$ MeV.

Following the work \cite{Fischer:2017lag}, the hadron part of the hybrid EoS with the quark phase described within the SFM we use the relativistic mean-field model with density dependent couplings DDf \cite{Alvarez-Castillo:2016oln}.
This hadronic EoS is relatively soft, which disables finding its Maxwell crossing with the stiff quark EoS of the $\chi$CDF
\footnote{The role of the stiffness and softness of quark and hadron EoSs in Maxwell construction is discussed in the review \cite{Baym:2017whm}.}.
Therefore, in this work the EoS of the $\chi$CDF was matched to the DD2 hadronic EoS \cite{Typel:2009sy}, which is stiffer than the DDf EoS \cite{Alvarez-Castillo:2016oln}.

\begin{figure}[!thb]
\begin{minipage}{\linewidth}
\includegraphics[width=0.9\columnwidth]{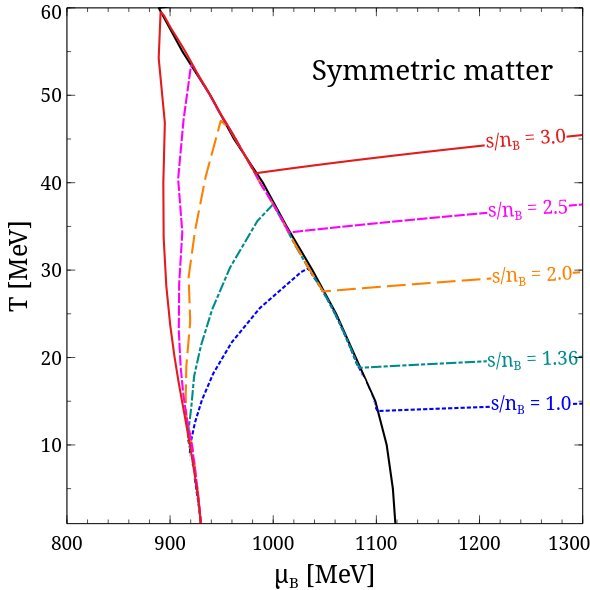}
\includegraphics[width=0.9\columnwidth]{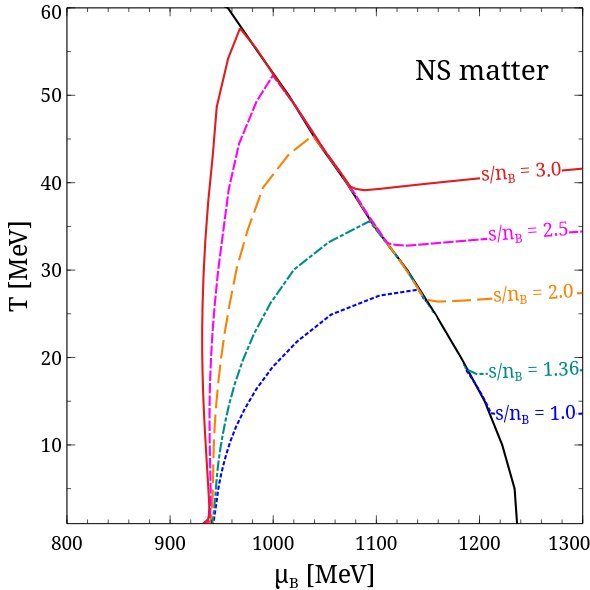}
\end{minipage}
\caption{Phase diagram of symmetric (upper panel) and NS (lower panel) matter in the plane of temperature $T$ versus baryon chemical potential $\mu_B$ obtained using the DDf-SFM hybrid EoS.
The black solid curves represent the phase boundary obtained according to the Maxwell construction.
The colored curves of different styles correspond to isontropes constructed for the values of entropy per baryon $s/n_B$.}
\label{fig1}
\end{figure}
Both DDf and DD2 EoS are relativistic density functionals for nuclear matter.
They reproduce the main properties of the ground state of nuclear matter and agree with
the low-density constraint of the chiral effective field
theory \cite{Kruger:2013kua}.
In the following, we refer to the two hybrid EoSs described above as DDf-SFM and DD2-$\chi$CDF.

\begin{figure}[t]
\begin{minipage}{\linewidth}
\includegraphics[width=0.9\columnwidth]{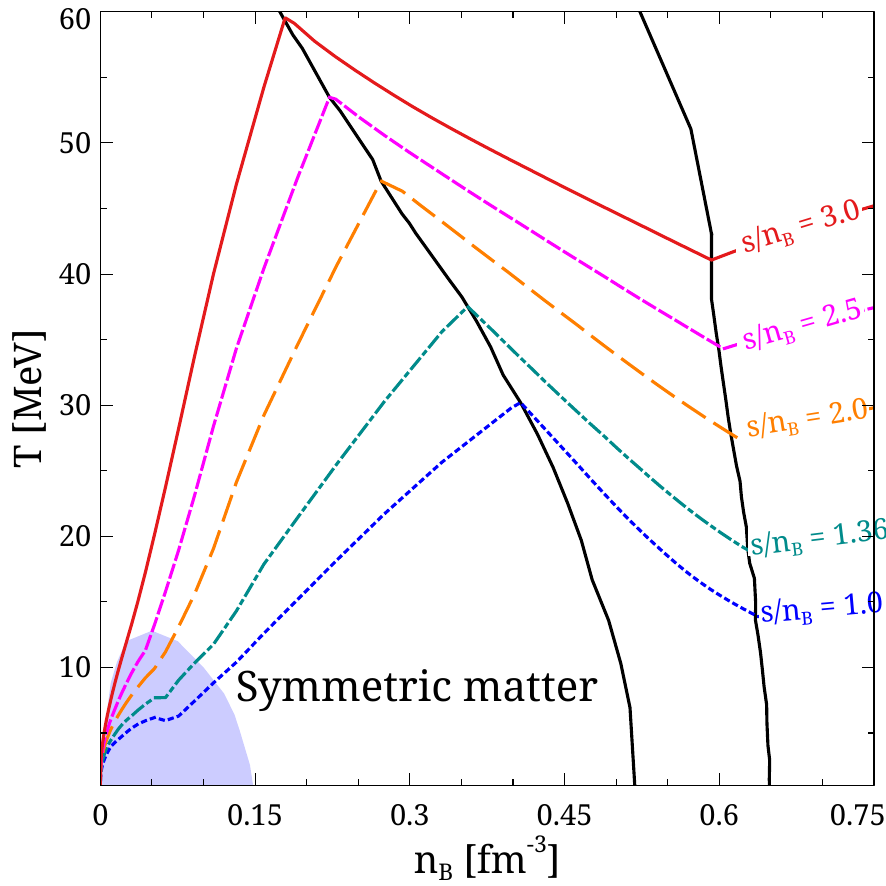}
\includegraphics[width=0.9\columnwidth]{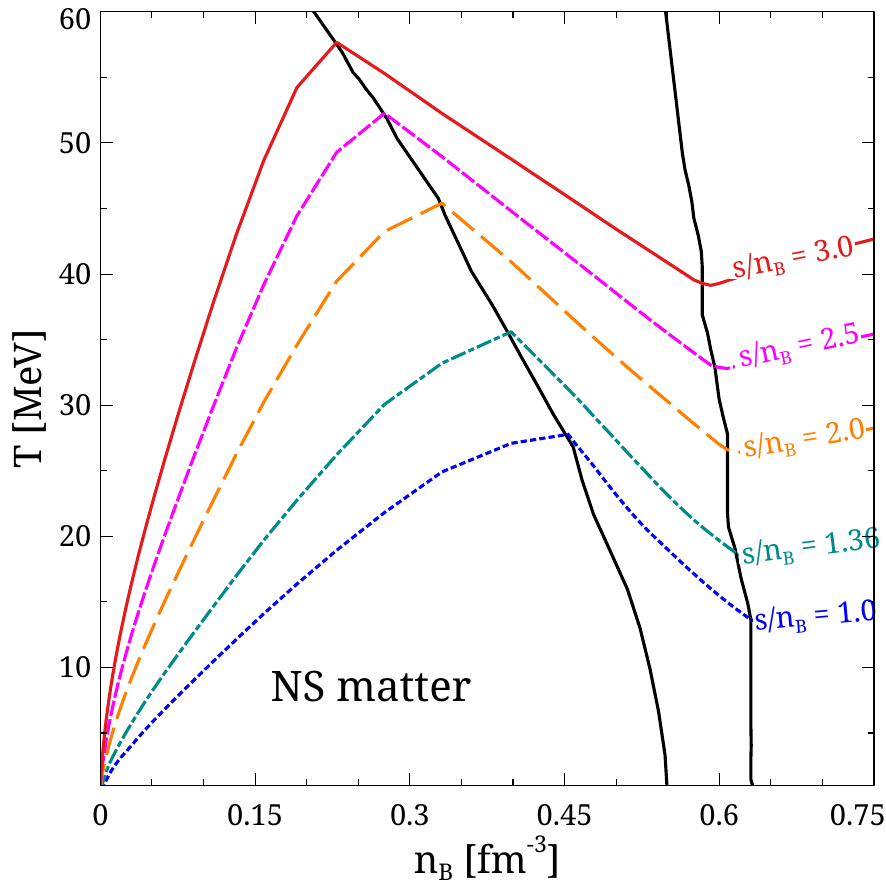}
\end{minipage}
\caption{The same as in Fig. \ref{fig1} but in the plane of temperature $T$ versus baryon density $n_B$.
The black solid curves represent the hadron and quark boundaries of the mixed phase obtained according to the Maxwell construction. The blue shaded area in the upper panel indicates the phase coexistence region of the gas-liquid transition in symmetric nuclear matter.}
\label{fig2}
\end{figure}

Fig. \ref{fig1} shows the phase diagram of the DDf-SFM EoS in the $T-\mu_B$ plane. 
As is seen, increasing temperature decreases the baryon chemical potential of quark deconfinement.
This happens because the response of quark matter to thermal excitations is stronger than that of hadronic matter. This, in turn, leads to a faster growth of the thermal pressure of quark matter compared to that of hadronic matter so that the line of critical baryon chemical potentials $\mu_B$ is bent to the left with increasing $T$.
The same conclusion can be drawn from Fig. \ref{fig2}, where the phase diagram of the DDf-SFM EoS is shown in the $T-n_B$ plane.
Both figures also show isentropes constructed for different constant values of entropy per baryon $s/n_B=$ 1.0 (blue), 1.36 (green), 2.0 (orange), 2.5 (magenta) and 3.0 (red). 
It is worth noticing that the temperature decreases when isentropes of the 
DDf-SFM go across the hadron-to-quark-matter phase transition. 
This is another indication of a stronger response of quark matter against thermal excitations when compared to that of hadronic matter. This behavior is a result of the increasing number of thermal degrees of freedom at the transition from hadronic to quark matter, which requires a lowering of the temperature to conserve the same entropy per baryon.

It was noted in \cite{Yudin:2015cva} that this decrease of the temperature along isentropes in the mixed phase causes the phenomenon of inverse convection which may have observable consequences for neutrino burst and gravitational wave signal in supernova explosions. This discussion was taken up in \cite{Jakobus:2022ucs}, where in Fig. 10 of that reference this behavior of the isentropes was denoted as entropic phase transition as opposed to an enthalpic transition when the temperature rises along isentropees in the hadron-to-quark matter phase transition. 

When comparing the two phase diagrams in Fig. \ref{fig2} for isospin symmetric matter (upper panel) and for neutron star matter (lower panel), it is astonishing why they are so similar.
They do not show the expected effect of a reduction in the onset density of deconfinement in the case of asymmetric matter. 
Because of the asymmetric occupation of the phase space of the isospin partners (neutrons and protons in the hadronic phase, up and down quarks in the deconfined phase), the critical chemical potential at the phase transition had to be higher in neutron star matter than in symmetric matter, see Fig. \ref{fig1}. 
This counter-intuitive behavior is due to the fact that in this parametrisation of the SFM model the isovector ($\varrho$- meson) meanfield in quark matter has been chosen such that the nuclear symmetry energy behaves continuous at the phase transition\footnote{
We want to point out that the choice of continuity of the symmetry energy that has been employed in the case of the SFM-DDf model is based on the argument that in the density functional approach to both nuclear and quark matter the symmetry energy originates from the isovector meson mean field so that the assumption of continuity throughout the phase border follows straightforwardly. 
Going beyond the mean field level, however, and accounting for the formation and dissociation of bound states and clusters, as well as a discontinuous behaviour of the quark mass as in the $\chi$CDF-DD2 model, the symmetry energy may exhibit a jump at the phase border. 
This entails that the onset density of deconfinement for the SFM-DDf model is basically the same for symmetric and NS matter while for the $\chi$CDF-DD2 model the discontinuity of the symmetry energy entails a lowering of the onset density for deconfinement in NS matter compared to the isospin-symmetric case.
}.  
Below, in the discussion of the results for the DD2-$\chi$CDF hybrid EoS, we will observe a different behavior because such an assumption of continuity of the symmetry energy has not been made. 
Without that assumption, the quark deconfinement in NS matter occurs at lower chemical potential and density than in isospin symmetric matter.

\begin{figure}[t]
\begin{minipage}{\linewidth}
\includegraphics[width=0.9\columnwidth]{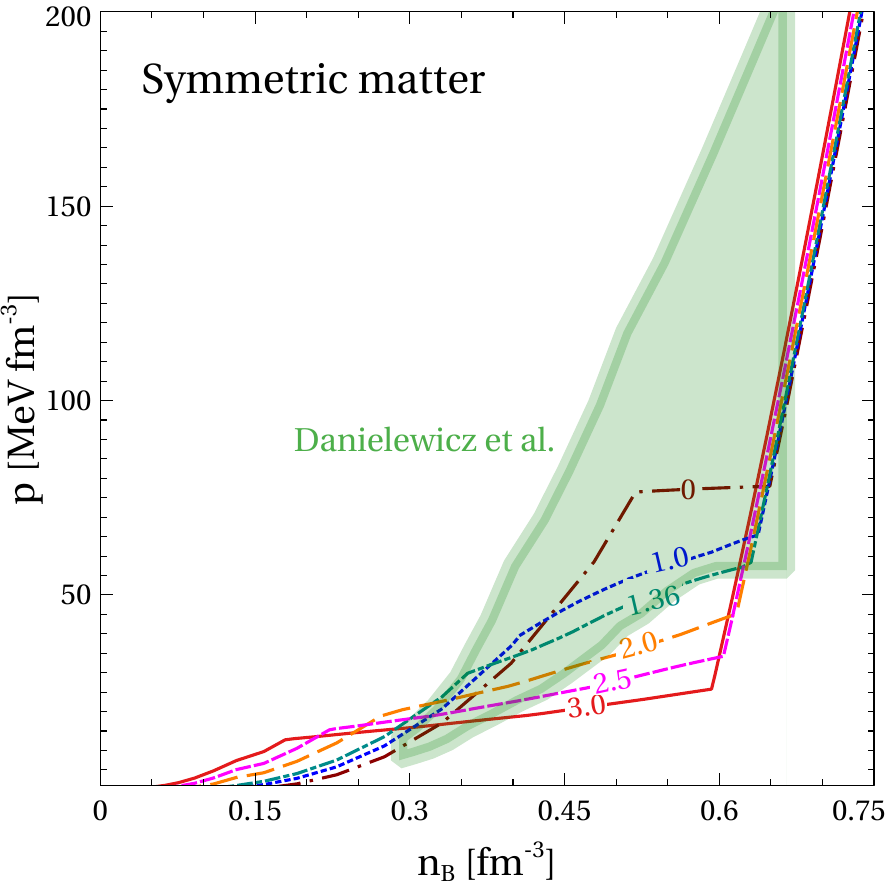}
\includegraphics[width=0.9\columnwidth]{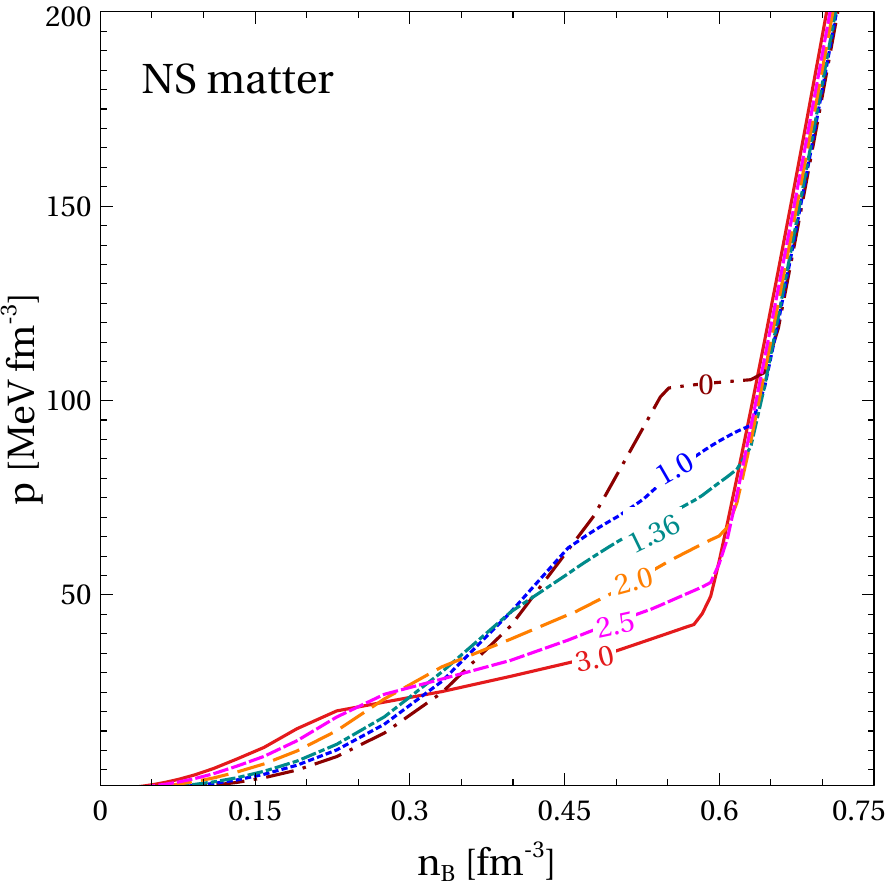}
\end{minipage}
\caption{Pressure $p$ as a function of baryon density $n_B$ of symmetric (upper panel) and NS (lower panel) matter of the DDf-SFM hybrid EoS.
The calculations are performed along the isentropes with the values of entropy per baryon $s/n_B$ indicated in the figure.
The shaded area in the upper panel corresponds to the proton flow constraint from Ref. \cite{Danielewicz:2002pu}.}
\label{fig3}
\end{figure}

Fig. \ref{fig3} demonstrates isentropic DDf-SFM EoS in the $p-n_B$ plane.
As is noticed earlier, increasing temperature or, equivalently, entropy per baryon leads to lowering the onset density of quark deconfinement.  
It is important that the soft character of the DDf hadronic EoS provides agreement of the DDf-SFM hybrid EoS with the proton flow constraint extracted from the experimental data on heavy ion collisions \cite{Danielewicz:2002pu}.
At the same time, increasing isospin asymmetry when switching from the regime of symmetric matter to the regime of NS matter stiffens the DDf-SFM hybrid EoS.
For example, in the cold matter case this corresponds to increasing the pressure of the mixed quark-hadron phase by about $25~{\rm MeV ~fm^{-3}}$.

\begin{figure}[t]
\begin{minipage}{\linewidth}
\includegraphics[width=0.9\columnwidth]{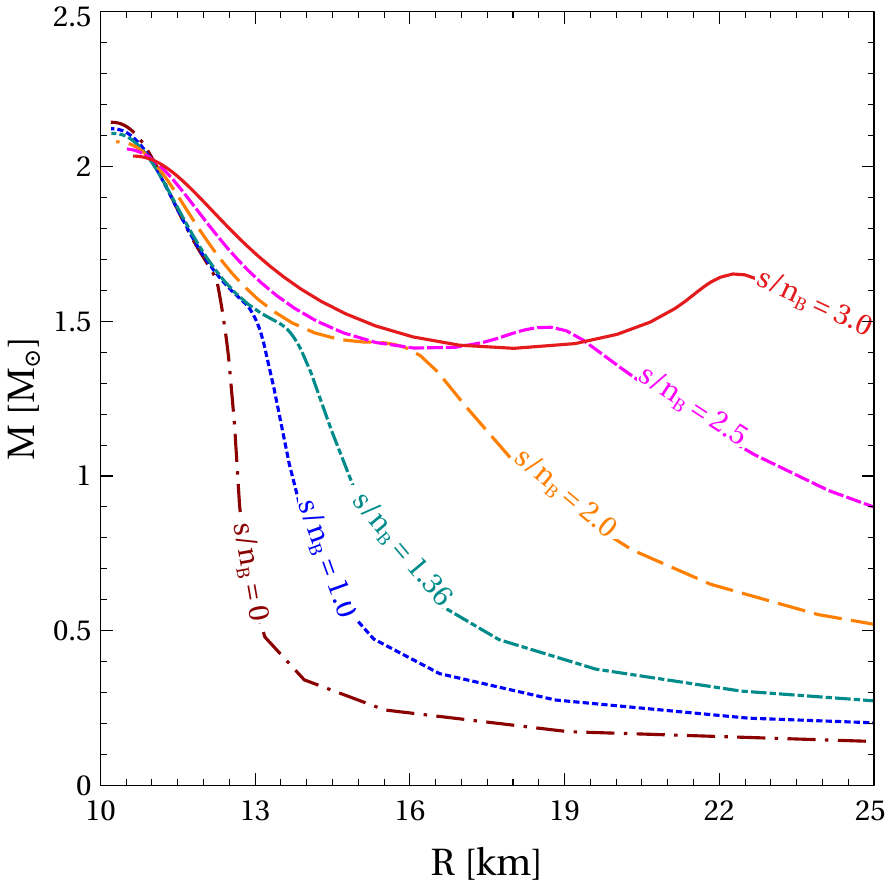}
\end{minipage}
\caption{Mass-radius relation of isentropic NS constructed with the DDf-SFM hybrid EoSs shown in the lower panel of Fig. \ref{fig3}.}
\label{fig4}
\end{figure}

Fig. \ref{fig3} demonstrates that the isentropic DDf-SFM EoS is characterized by decent growth of the density jump across the phase transition when $s/n_B$ is increasing, while the high density quark part of this EoS is rather insensitive to $s/n_B$.
It is also seen that isentropes with higher entropy per baryon are located below the isentropes with smaller values of $s/n_B$.
This behaviour corresponds to a softening of the finite-temperature EoS which, e.g., in heavy-ion collisions may help to improve the description of kaon production, see the discussion in \cite{Klahn:2006ir} and \cite{Fuchs:2005zg}. 

At sufficiently high entropy per baryon the density jump reaches the value leading to gravitational instability of NSs due to strong first order phase transition in their interiors \cite{1971SvA15347S}.
If followed by significant stiffness of the high density phase, such instability leads to formation of twin star configurations of NSs.
In this case in a narrow range of masses pairs of the NSs with the same masses have different radii \cite{Blaschke:2019tbh,Goncalves:2022phg,Pradhan:2023zmg,Chanlaridis:2024rov,Li:2024sft,Sabatucci:2026qcz}. 
The large radius counterparts of such twin NS pairs are purely hadronic, while the small radius NSs of the pairs contain quark cores.
Thus, the DDf-SFM EoS has the features of thermally induced twin NSs.

This conclusion is confirmed by Fig. \ref{fig4}, which shows mass-radius relations of the isentropic hybrid DDf-SFM EoS. 
At a value between $s/n_{B}= 1.36 ... 2.0$, the mass-radius mapping loses its uniqueness and for $s/n_{B}\ge 2.0$ we see that in a mass range around $1.5~M_\odot$ one finds two configurations with different radii for the same mass, the mass twins.
We can estimate the critical value of the entropy per baryon where this occurs for the first time by applying the Seidov criterion \eqref{eq:seidov}, now in a form for the density jump at the transition; 
\begin{equation}
    \label{eq:seidov2}
    \Delta n \ge n_c \left(\frac{1}{2} + \frac{1}{1+\varepsilon_c/p_c}  \right) \approx \frac{3}{4} n_c ~~,~ {\rm for~} \varepsilon_c\approx 3 p_c.
\end{equation}

Reading off from Fig. \ref{fig3} that for  $s/n_{B}=1.36~(2.0)$ the onset density is $n_c=0.4~(0.32)$ and $\Delta n=0.24~(0.28)$, one obtains 
$\Delta n - 3n_c/4=-0.06~(+0.04)$. By simple linear interpolation the critical value for the onset of the twin instability is $s/n_{B}=1.75$. 
Comparing with Fig. \ref{fig4}, one concludes that the estimate with the Seidov criterion \eqref{eq:seidov2} works well despite the fact that it was derived for the zero temperature case.
A similar conclusion was drawn by the authors of \cite{Carlomagno:2024vvr},
where for hybrid EoS based on a nonlocal chiral quark model with a density dependent bag pressure the thermal twin onset was obtained for $s/n_{B}=1.81$, i.e. rather close to the present result.

\begin{figure}[!thb]
\begin{minipage}{\linewidth}
\includegraphics[width=0.9\columnwidth]{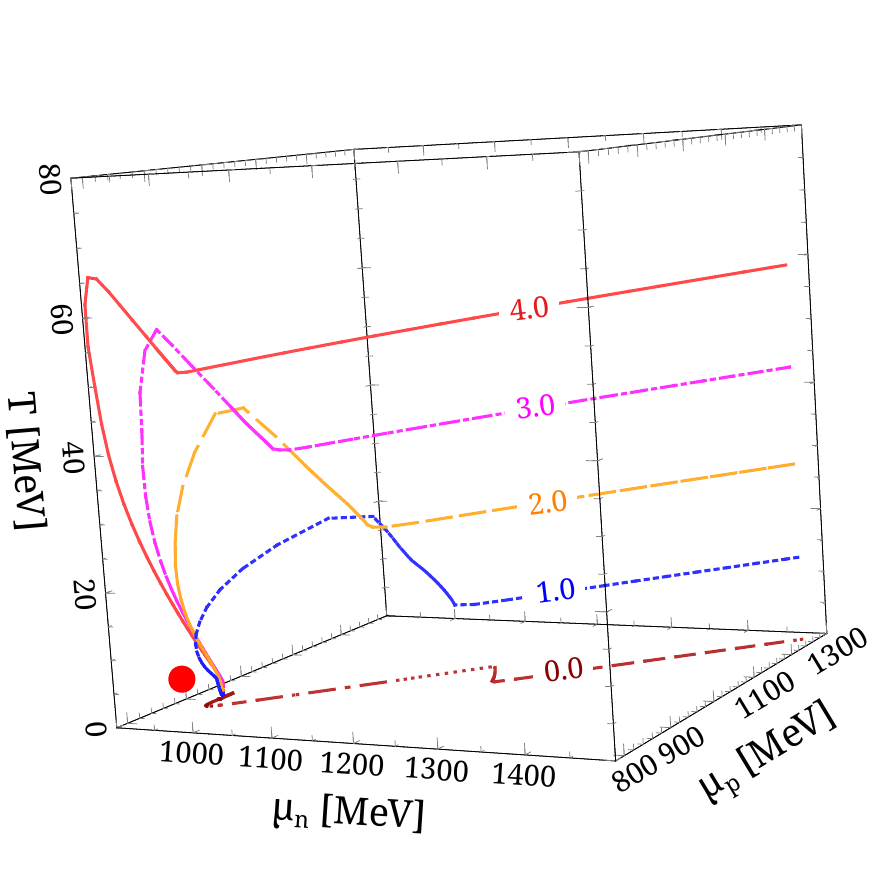}
\end{minipage}
\caption{Trajectories of constant entropy per baryon $s/n_B$ indicated in the figure in the three-dimensional phase diagram of temperature $T$ and chemical potentials for protons ($\mu_p$) and neutrons ($\mu_n$) under conditions of electric neutrality and beta-equilibrium with leptons.
The red dot shows the location of the freeze-out parameters of a nuclear matter droplet which would fit the distribution of heavy r-process elements, see \cite{Ropke:2025idq,Blaschke:2025}.}
\label{fig5}
\end{figure}

Before we turn to the discussion of the DD2-$\chi$CDF hybrid EoS, we want to show in Fig. \ref{fig5} the evolution along constant entropy per baryon trajectories in the three-\- dimensional phase diagram of temperature $T$ and chemical potentials for protons ($\mu_p$) and neutrons ($\mu_n$) under conditions of electric neutrality and beta-equilibrium with leptons.
The magenta trajectory for $s/n_B=1$ comes very close to the red dot, which shows the location of the freeze-out parameters of a nuclear matter droplet which would fit the quantum statistical distribution of heavy r-process elements. For more details of the heavy element freeze-out, see \cite{Ropke:2025idq,Blaschke:2025} and for the embedding of such a scenario into the evolution of the early Universe as a byproduct of the era of primordial black-hole formation in the cosmic QCD transition, see 
\cite{Gonin:2025uvc,Gonin:2026xhe}.

\begin{figure}[t]
\begin{minipage}{\linewidth}
\includegraphics[width=0.9\columnwidth]{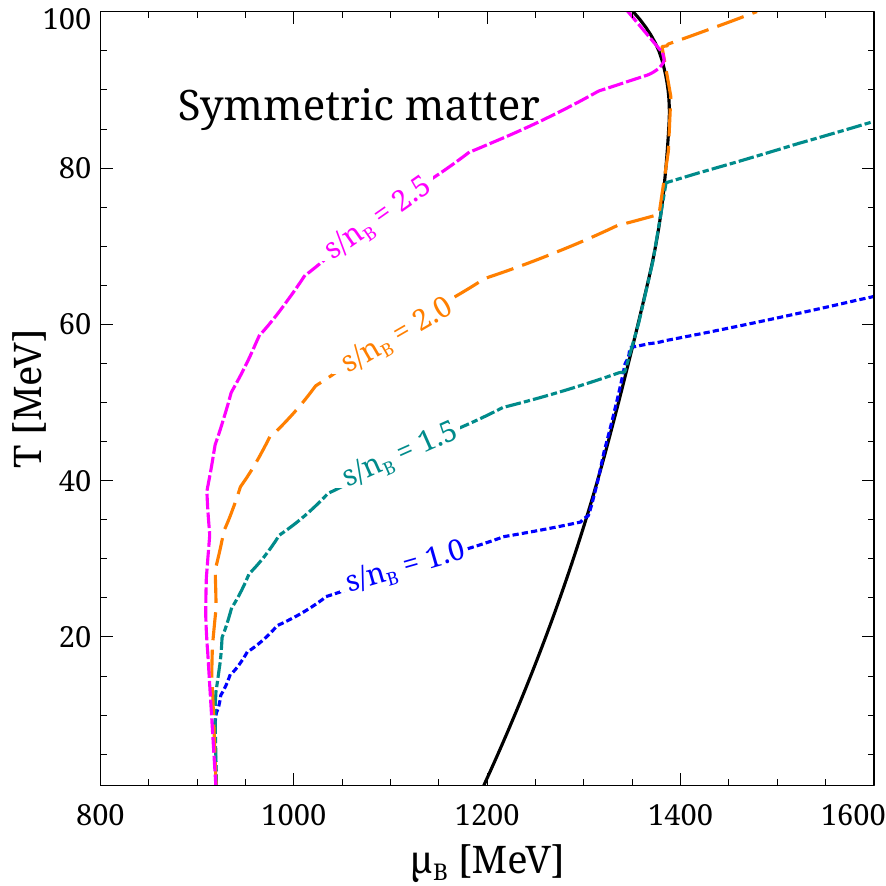}
\includegraphics[width=0.9\columnwidth]{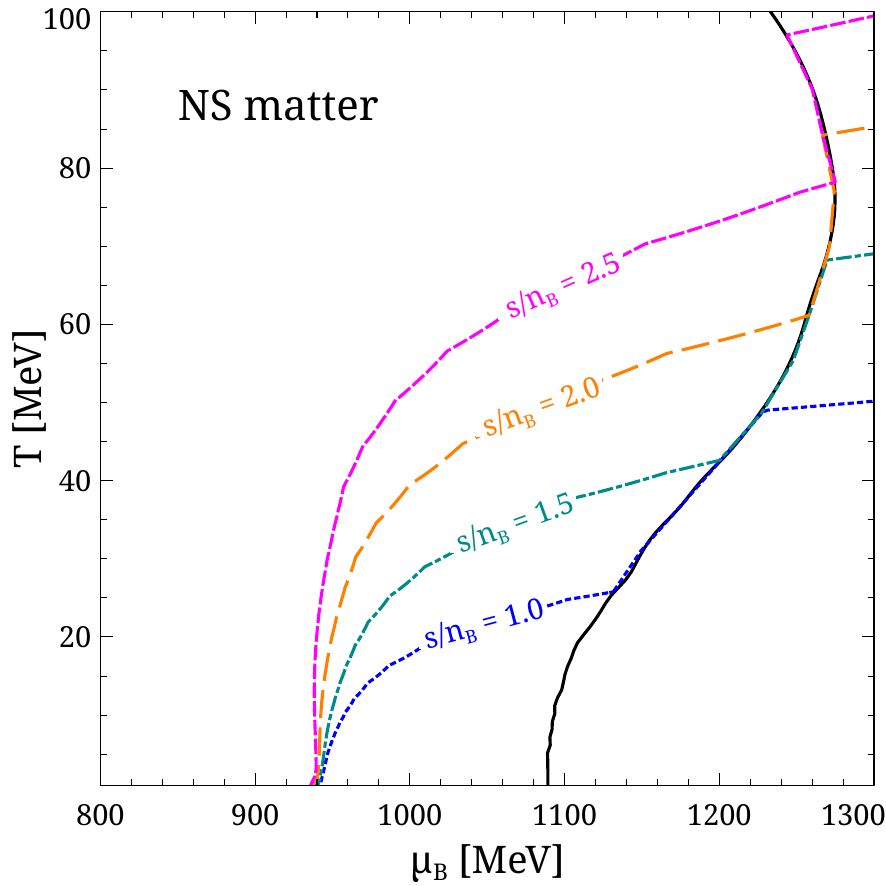}
\end{minipage}
\caption{The same as in Fig. \ref{fig1} but for the DD2-$\chi$CDF hybrid EoS.}
\label{fig6}
\end{figure}
\begin{figure}[!htb]
\begin{minipage}{\linewidth}
\includegraphics[width=0.9\columnwidth]{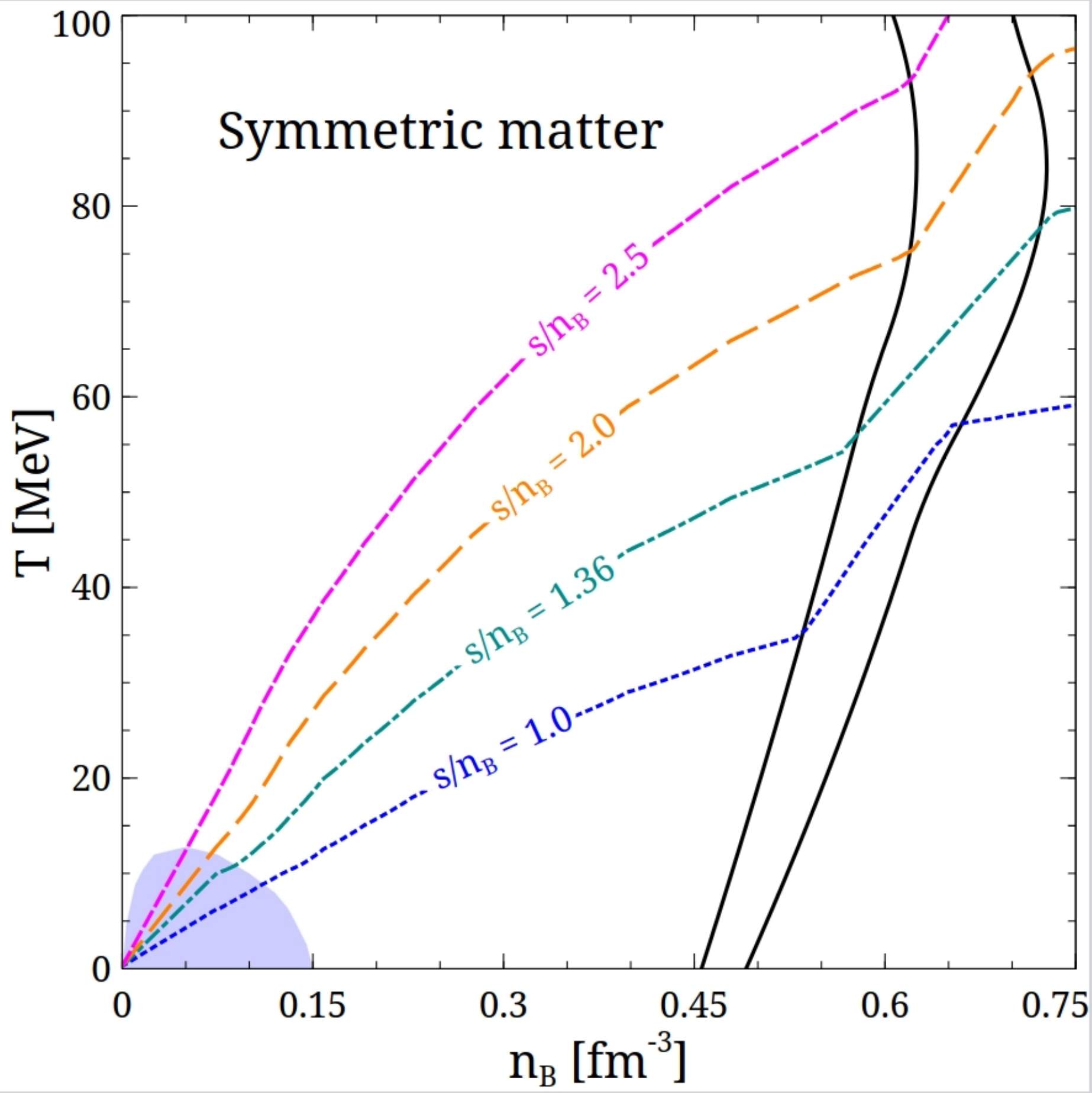}
\includegraphics[width=0.9\columnwidth]{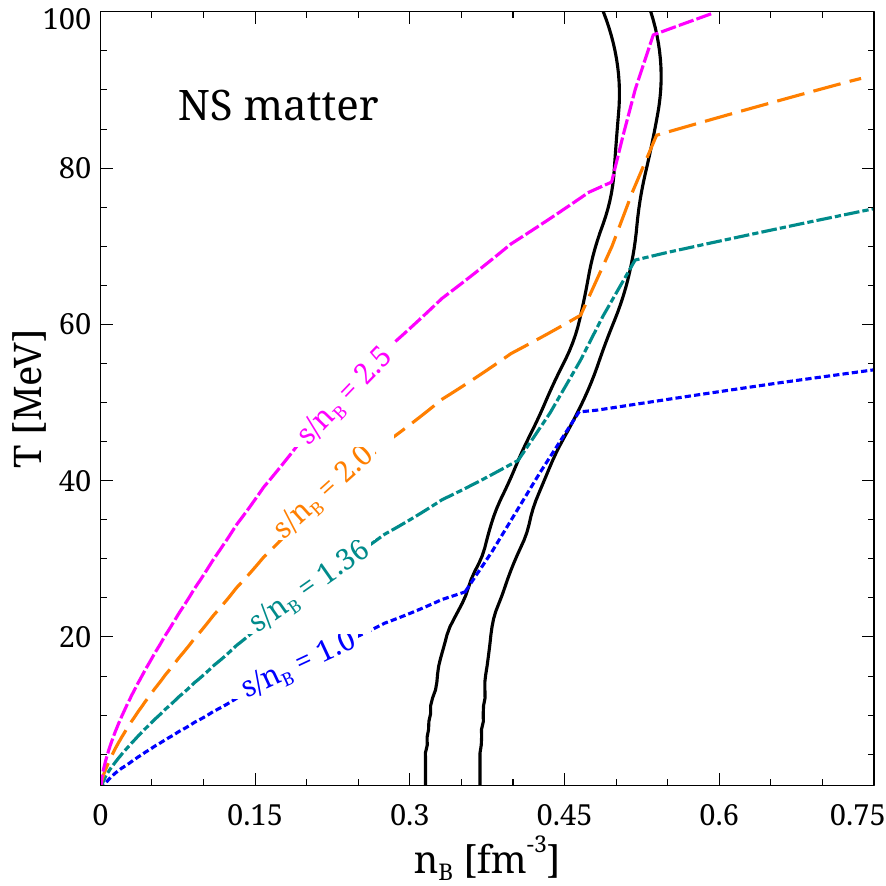}
\end{minipage}
\caption{The same as in Fig. \ref{fig2} but for the DD2-$\chi$CDF hybrid EoS.}
\label{fig7}
\end{figure}

Figs. \ref{fig6} and \ref{fig7} show the phase diagram of the DD2-$\chi$CDF EoS in the $T-\mu_B$ and $T-n_B$ planes, respectively.
The main qualitative difference of this phase diagram compared to the phase diagram of the DDf-SFM EoS is that the corresponding phase boundary is bent rightward so that the chemical potential and density of the onset of quark deconfinement grow with increasing temperature.
This is a direct consequence of 2SC in quark matter.
In the case of 2CS diquark correlations of quarks are relevant degrees of freedom.
The flavor and color states of the quarks that make up diquarks are strongly entangled \cite{Buballa:2003qv}. 
This leads to a significant reduction of the number of effective degrees of freedom in paired quark matter.
Therefore, unlike the unpaired case of the DDf-SFM EoS, the growth of pressure due to thermal excitation in quark matter is smaller than the corresponding growth in hadron matter and fulfilling the Maxwell criterion of phase equilibrium requires higher densities, i.e. bending the phase boundary rightward.  

\begin{figure}[!t]
\begin{minipage}{\linewidth}
\includegraphics[width=0.9\columnwidth]{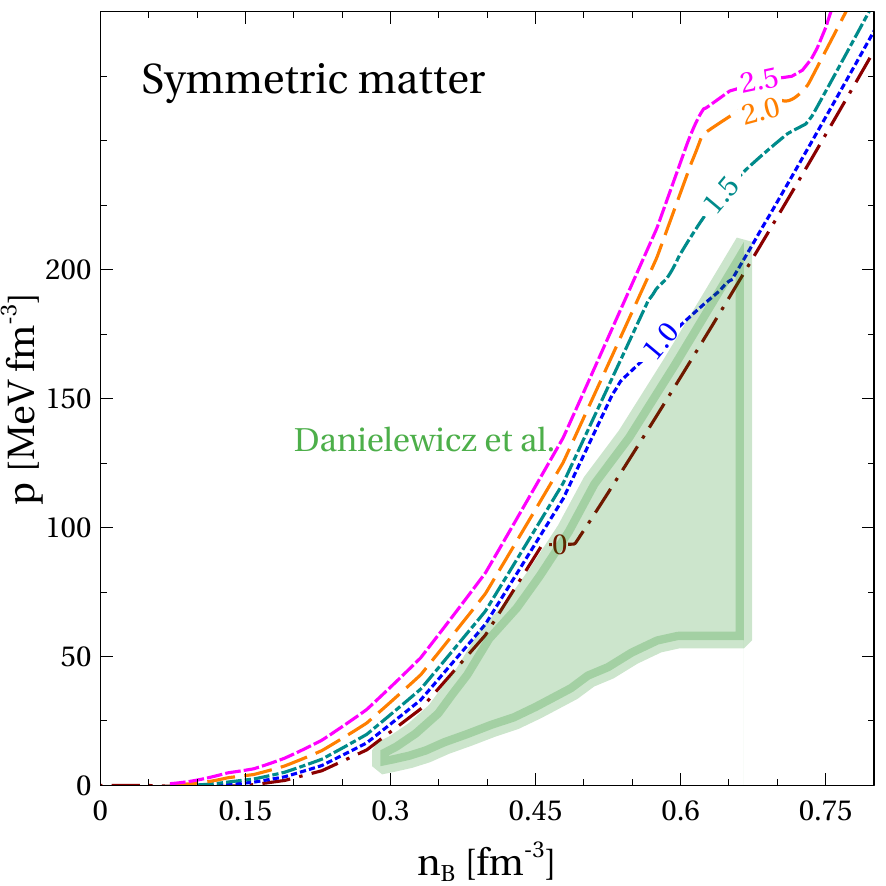}
\includegraphics[width=0.9\columnwidth]{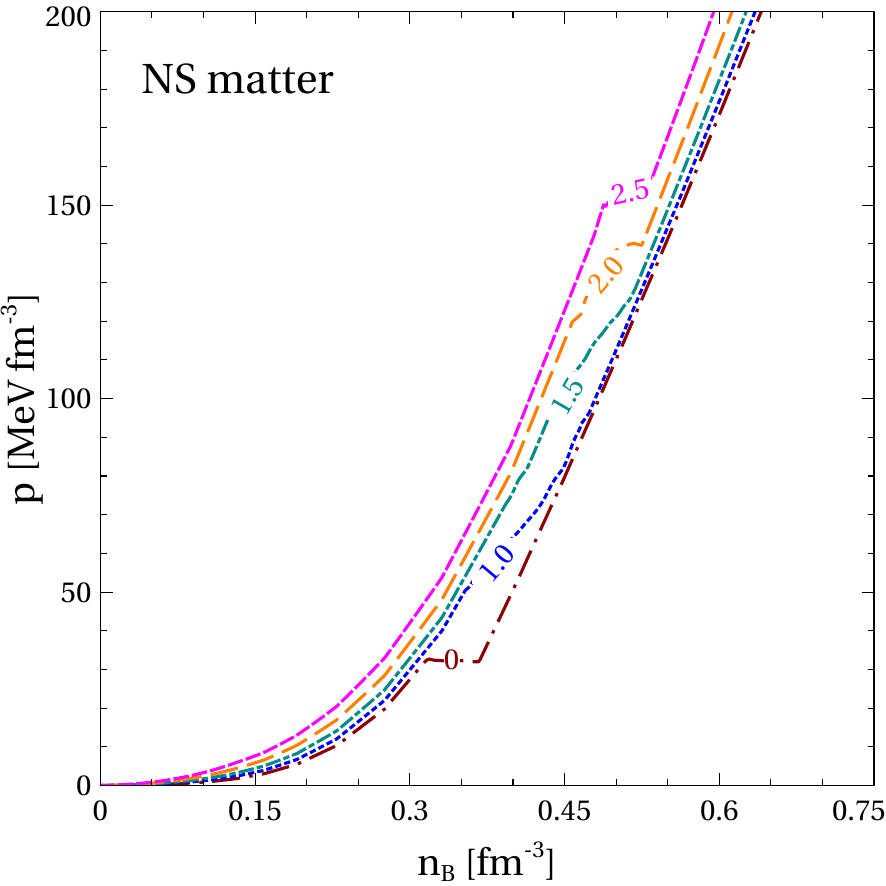}
\end{minipage}
\caption{The same as in Fig. \ref{fig3} but for the DD2-$\chi$CDF EoS.}
\label{fig8}
\end{figure}

Another effect of 2SC corresponds to temperature growth across the deconfinement phase transition, which is seen from the behavior of isentropes shown in Figs. \ref{fig6} and \ref{fig7}.
To explain it, we notice that in the considered case maintaining a certain value of entropy per baryon requires higher temperatures since the effective number of degrees of freedom reduces due to the entanglement of quark flavor and color discussed above.
Such effects have already been observed (see, e.g. Ref. \cite{Ivanytskyi:2022wln}).

The density jump at the deconfinement phase transition in the DD2-$\chi$CDF EoS is significantly smaller than in the DDf-SFM case.
This can be seen from Fig. \ref{fig8} showing the isentropic DD2-$\chi$CDF EoS. 
At the same time, contrary to the non-color-superconducting case of the DDf-SFM EoS, the 2SC phase of quark matter is considerably affected by temperature.
This is caused by the response of paired quark matter to melting the diquark condensate at temperature growth. 
It is also worth mentioning that in the presence of 2SC pairing, increasing $s/n_B$ leads to moving the isentrope upward, which is the opposite behavior compared to the non-color-superconducting DDf-SFM EoS.

\begin{figure}[!                                                                                     htb]
    \centering
    \includegraphics[width=0.9\columnwidth]{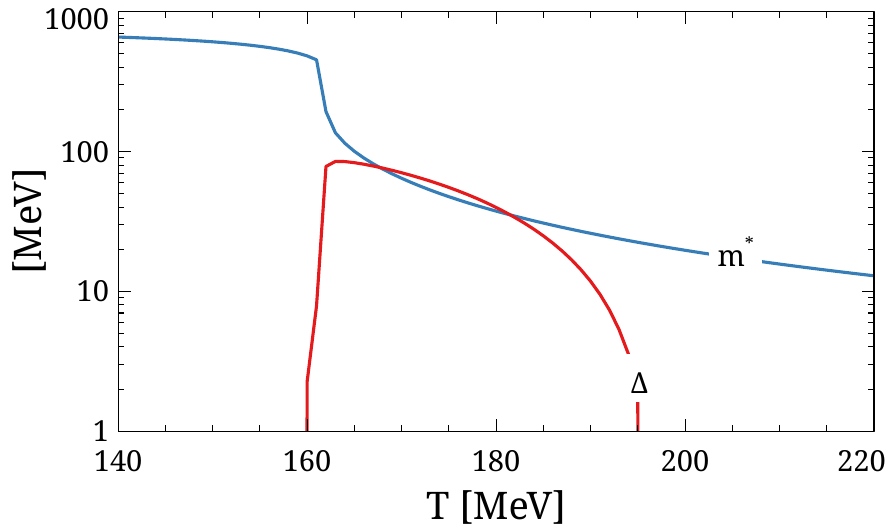}
    \includegraphics[width=0.9\columnwidth]{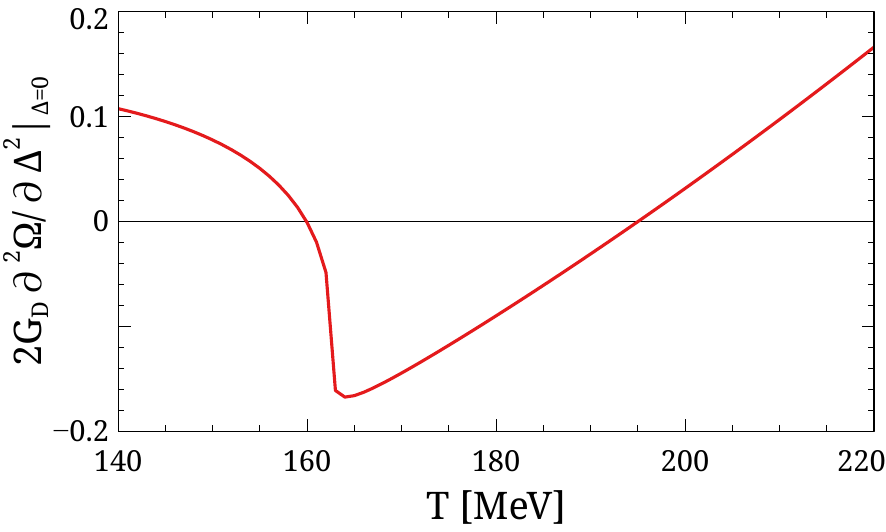}
    \caption{Upper panel: temperature dependence of the dynamical quark mass $m^*$ (blue solid line) and the diquark pairing gap $\Delta$ (red solid line) at vanishing baryochemical potential $\mu_B=0$. Lower panel: criterion for the occurrence of a color superconducting phase. For details, see text.    
    }
    \label{fig9}
\end{figure}

The question arises for the critical temperature of the transition from 2SC superconducting quark matter to normal quark matter. To answer this question, we consider the criterion \cite{Ivanytskyi:2022oxv}
\begin{equation}
\label{eq:Tc}
    \frac{\partial^2 \Omega}{\partial \Delta^2}\bigg|_{\Delta=0} = 0
\end{equation}
and solve the corresponding equation for the critical temperature at $\mu_B=0$.
The result is depicted in the lower panel of Fig. \ref{fig9}, which exhibits two zeroes for the appearance and disappearance of color superconductivity at $T=160$ MeV and $T=195$ MeV, respectively.

Due to strong color superconductivity in the $\chi$CDF quark matter phase, there is a qualitatively new effect, illustrated in the upper panel of Fig. \ref{fig9}, where the dynamical quark mass $m^*$ and the diquark pairing gap $\Delta$ are shown as functions of the temperature for $\mu_B=0$.

Thus, a characteristic, unique feature of the  $\chi$CDF approach becomes apparent:
the hadronic phase has no direct border with the normal quark matter phase, but rather with the 2SC color superconducting phase, even at $\mu_B=0$.
We note that both the quark mass and the diquark gap do not obey the BCS relation between (pseudo-)critical temperature and gap at zero temperature. 

The dramatic drop in the quark mass at $T_c\sim 160$ MeV is indicating the partial chiral restoration which induces the Mott dissociation of hadronic bound states when their masses hit the continuum edge for unbound quark states which delimits the hadronic matter phase. The nonvanishing diquark pairing gap in the temperature range $160 < T[{\rm MeV}] < 195$ signals an intermediate color superconducting phase between hadronic and normal quark matter as a unique feature of the $\chi$CDF model.

\begin{figure*}[!htb]
    \includegraphics[width=0.9\columnwidth]{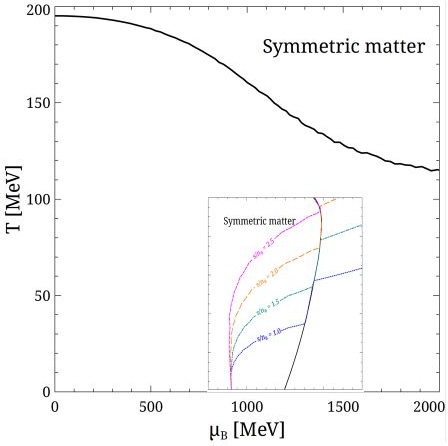}
\includegraphics[width=0.9\columnwidth]{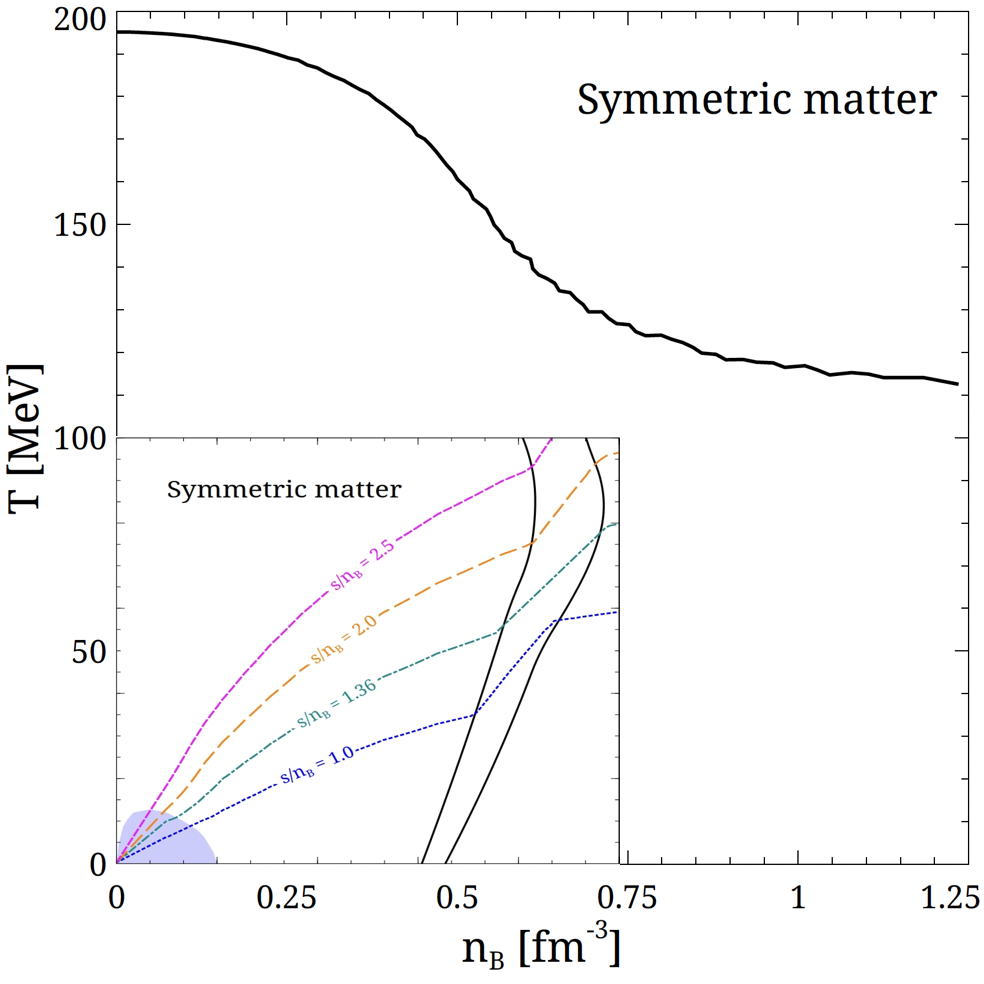}
\caption{Critical temperature for 2SC color superconductivity vs. baryon chemical potential (left panel) and vs. baryon density (right panel) for isospin-symmetric matter.
In the insets the corresponding parts of the phase diagrams of Figs.  \ref{fig6} and \ref{fig7} are shown. It is a characteristic, unique feature of the DD2-$\chi$CDF model that the hadronic phase has no direct border with the normal quark matter phase, but rather with the 2SC color superconducting phase, even at $\mu_B=0$, see Fig. \ref{fig9}.}
\label{fig10}
\end{figure*}

We have solved the criterion \eqref{eq:Tc} for the critical temperature of color superconductivity at arbitrary $\mu_B$ and show the result in the phase diagrams of the DD2-$\chi$CDF model (see Fig. \ref{fig10}), where as insets the phase borders between hadronic and 2SC quark matter at moderate temperatures from Figs. \ref{fig6} and \ref{fig7} are displayed. 
In this figure, the remarkable, unique feature of the DD2-$\chi$CDF model becomes apparent, that between the hadronic phase and the normal quark matter phase, there is a corridor of  2SC color superconducting quark matter.

\begin{figure}[!htb]
\begin{minipage}{\linewidth}
\includegraphics[width=\columnwidth]{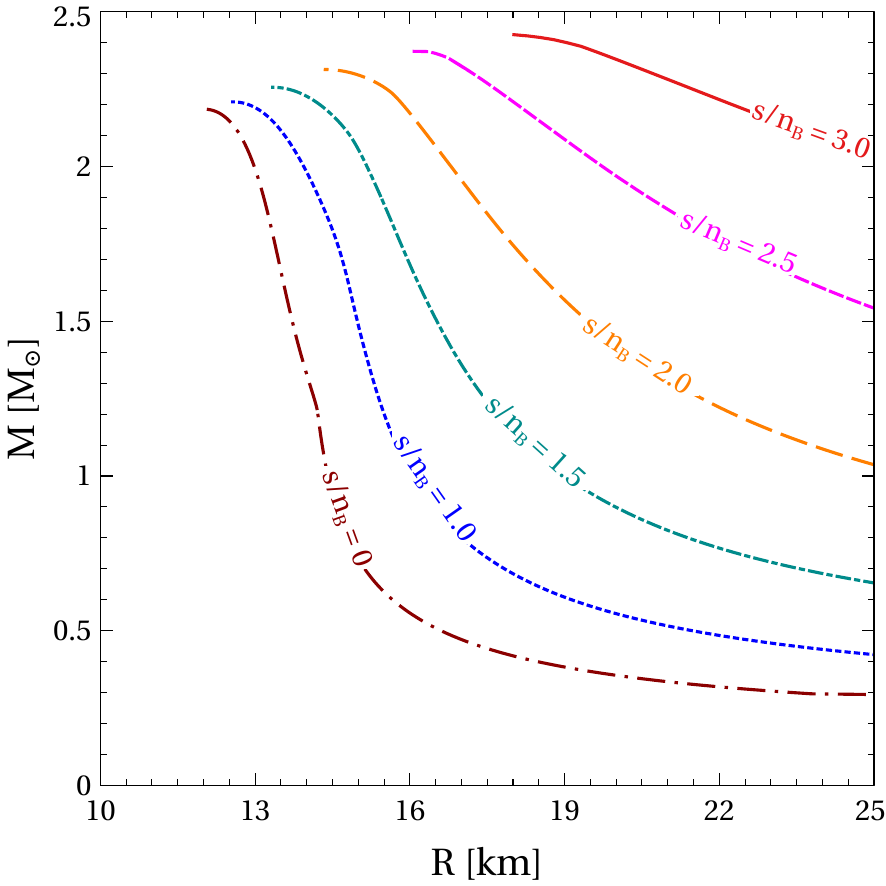}
\end{minipage}
\caption{The same as in Fig. \ref{fig4} but for the DD2-$\chi$CDF hybrid EoSs shown in the lower panel of Fig. \ref{fig8}.}
\label{fig11}
\end{figure}

In Fig. \ref{fig11}, we show the result of the integration of the TOV equations for EoS with $s/n_B=$ const for the DD2-$\chi$CDF hybrid EoS.
As to be expected from the systematics of the NS matter EoS in Fig. \ref{fig8} which exhibits a stiffening with increasing values of $s/n_B$,
we see no gravitational instability that could lead to thermal twin stars.
Due to the stiffening of the EoS, both the masses and radii are increased with increasing temperature and the curves form an onion-shaped family of star sequences. There is no clear structure related to the onset of deconfinement and in particular the feature of thermal twin stars is absent. 
%
\section{Summary and Discussion}
\label{sec_summary}

In this work we have discussed the hybrid EoS for quark-nuclear matter within a relativistic density functional model that was obtained by a Maxwell construction using the standard relativistic density functionals DD2 and DDf for nuclear matter and the newly developed confining density functionals for quark matter in the normal (SFM) and in the color superconducting ($\chi$CDF) phase. 
We compared the isospin-symmetric matter case relevant for heavy-ion collisions with the neutron star matter case and observe a lowering of the onset of deconfinement due to the asymmetrically occupied phase space in the NSM case. However, when the symmetry energy is adjusted so that it is continuous at the phase border (DDf-SFM case), then the phase diagrams are almost identical.

When considering strong color superconductivity, there are dramatic effects on the phase diagrams and the EoS:
\begin{enumerate}
\item Diquark condensation lowers the onset of chiral symmetry restoration and thus also the onset of deconfinement;
\item Strong color superconductivity alters the shape of the phase border at low temperatures so that with increasing temperature the critical density rises;
\item Constant $s/n_B$ curves correspond to rising (dropping) temperatures in the mixed phase with (without) strong diquark condensation;
\item The hadronic phase has no common border with normal quark matter in the phase diagram, so that deconfinement proceeds to color superconducting matter, even at zero baryon density;
\item No thermal twin stars with strong color superconductivity and possibly therefore no explodability of quark deconfinement supernovae.
\end{enumerate}

These dramatic effects are a consequence of sufficiently strong diquark coupling. They are not specific to the CDF models of quark matter and are not related to the confining effect of enhanced chiral symmetry breaking by a medium-dependent scalar meson coupling  
\footnote{We like to prove this statement by giving examples where the above effects have been seen in NJL-type models: 1) the lowering of the chiral symmetry restoration density with increasing diquark coupling is demonstrated in \cite{Blaschke:2005uj,Ruester:2005jc}, 2) the backbending of the phase border in the low-temperature phase diagram has been shown in \cite{Blaschke:2010ka}, where also 3) the rising (dropping) temperature for 
$s/n_B={\rm const}$ curves in the mixed phase for large (small) pairing gaps is shown; see \cite{Kunkel:2026wvz} for a recent disussion of isentropes in the phase diagram from the renormalization-group consistent NJL model; 4) the trend that with increasing diquark coupling the 2SC phase approaches the temperature axis is shown in \cite{Zablocki:2009ds} and the absence of CCSN explosion for the color superconducting three-flavor NJL model has been discussed in \cite{Fischer:2011zj}.}. 
As has been shown by Ivanytskyi and Blaschke \cite{Ivanytskyi:2022oxv}, in the restored chiral symmetry phase where color superconductivity occurs, the scalar and pseudoscalar couplings are degenerate and become the same as in the NJL model case.
Therefore, one can conclude that for the CDF models, there must also be a sufficiently weak diquark coupling that still allows a color superconducting phase but without the dramatic effects on the QCD phase diagram enumerated above. Consequently, thermal twins and therefore CCSN explodability by the QCD deconfinement mechanism must be possible also for color superconducting CDF models. The existence of such a moderate diquark coupling is a new result of the present work. Its value must be quantified in future research.

The advantage of CDF models over the local \cite{Blaschke:2010ka,Sabatucci:2026qcz} or nonlocal \cite{Carlomagno:2023nrc,Carlomagno:2024vvr} NJL models is that the former do not require the ad hoc adjustment of a constant or medium dependent bag pressure to allow for the possibility of reasonable phase transitions with the Maxwell construction.
 
In order to remove arbitrariness and to sharpen predictions of the QCD phase structure at low temperatures and high baryon densities, the description of baryons as bound states of quarks and baryonic matter on this basis should be further developed. Interesting first steps were presented in \cite{Bentz:2001vc,Wang:2010iu}.

While here (and in most other models using confining density functionals) the coupling constant of the confining density functional is considered as a free parameter with the dimension of the string tension, one may compare the range of effective couplings with results that are derived from a confining interaction potential fitted to the spectroscopy of heavy quarkonia for different schemes of color saturation (screening) \cite{Heymer:2026sox}.
Such a comparison may provide insights to the microphysics of many-quark systems with confining interactions and their macroscopic manifestations.

\section{Conclusions and Outlook}
\label{sec_conclusion}

In this work, we demonstrated that the CDF approach to quark matter can eliminate certain caveats of other approaches to the quark matter phase in the QCD phase diagram and in applications to heavy-ion collisions, neutron stars, neutron star mergers and core-collapse supernovae.
The local and nonlocal NJL-type models, for example, have to assume ad hoc bag pressures to facilitate a phase transition, sometimes even in a medium-dependent
manner, dropping with density \cite{Alvarez-Castillo:2018pve,Carlomagno:2023nrc} or just being negative so that the onset density gets lowered \cite{Sabatucci:2026qcz}.

As it has been exercised in ref. \cite{Alvarez-Castillo:2018pve}, the medium-dependent bag pressure (and vector coupling) can be adjusted such as to reproduce the SFM thermodynamics within the setting of a generalized nonlocal NJL model, where density dependent coupling and bag pressure parameters are adopted. Such ad hoc assumptions are not satisfactory. 

The features of the QCD phase diagram explained in this work are consequences of the thermodynamic properties of the EoS model, not specific to the underlying microphysical approach. However, the CDF approach presented here has a distinct advantage over NJL type models as it does not require the introduction of an ad hoc bag pressure.

We want to point out a strategy for the further development of the $\chi$CDF approach.

In this work, we have applied a simple Maxwell construction to obtain a hybrid quark-hadron EoS. 
On the basis of arguments which led to quark-hadron continuity at the deconfinement transition, and because the properties of hadronic and quark matter in the region of the deconfinement transition are not well known, constructions of a crossover EoS have been applied
\cite{Masuda:2012ed,Albright:2014gva,Kapusta:2021ney,Blaschke:2021poc,Ayriyan:2021prr,Ivanytskyi:2022wln}.
Going beyond such constructions, which have to assume so-called "switch functions", there are concepts based on the generalized Beth-Uhlenbeck approach \cite{Blaschke:2023pqd} which could be developed in order to predict the phase structure in the domain of densities and temperatures that is not accessible to ab-initio Lattice QCD simulations. 
This is a wide field for future work.

In such a scheme, the hadronic matter phase appears as a clustered quark matter phase, where hadrons are described as bound states of quarks, thus going beyond the mean-field approximation.

A cluster decomposition of the $\Phi$ functional in the field-theoretic setting of the generalized Beth-Uhlenbeck approach suggested in Ref. \cite{Blaschke:2023pqd} shall be carried out, including the relevant hadronic correlation channels with their bound and scattering state spectra. 
As it has been demonstrated in \cite{Bastian:2018mmc} for a strongly schematized setting, in such an approach it is possible to predict the location of a CEP in the QCD phase diagram. This CEP may be located at rather low temperatures or be absent altogether, so that a crossover-all-over situation would emerge.
But even in a crossover-type situation it is not excluded that thermal twin stars would emerge which according to the bold conjecture of this work, may be indicators of the CCSN explodability of the corresponding EoS, as required by astronomical observations.

Concluding this work, we have suggested that the presence of thermal twin stars as indicator of CCSN explodability of massive progenitor stars may be considered as criterion of the "validity" of a QCD phase diagram model. 
Future investigations have to show whether this criterion may be compatible with the presence of a weaker color superconducting phase than has been elaborated in this work. 


\subsection*{Data availability statement}

The data produced by this article are publicly available on the ZENODO repository at 
https://zenodo.org/records/21613117 .

\subsection*{Acknowledgement}
We thank Tobias Fischer for his collaboration in an early phase of this work and numerous discussions of the QCD deconfinement supernova explosion mechanism. Pasi Huovinen is acknowledged for his critical remarks and insightful comments on the hydrodynamic description of heavy-ion collisions. We are grateful to Andrey Yudin and Selina Kunkel for pointing out to us the references \cite{Yudin:2015cva} and \cite{Kunkel:2026wvz}, respectively. 
This research is part of the project No. 2021/43/P/ST2/03319 co-funded by the
National Science Centre and the European Union Framework Programme for Research and Innovation Horizon 2020 under the Marie Skłodowska-Curie grant agreement No. 945339. For the purpose of Open Access, the author has applied a CC-BY public copyright licence to any Author Accepted Manuscript (AAM) version arising from this submission.
O.I. acknowledges support from the Scultetus visiting scientist program for his visit at HZDR/CASUS G\"orlitz, where this work has been completed.

\bibliography{bibliography}

\end{document}